\documentclass{emulateapj}

\newcommand{\newton}{X-ray Multi-Mirror Mission (XMM-Newton)}
\newcommand{\newtonshort}{XMM-Newton}
\newcommand{\xmm}{XMM-Newton}
\newcommand{\chandra}{Chandra}
\newcommand{\rxte}{RXTE}
\newcommand{\swift}{Swift}
\newcommand{\suzaku}{Suzaku}
\newcommand{\nustar}{NuSTAR}
\newcommand{\astrosat}{Astrosat}
\newcommand{\integral}{INTEGRAL}
\newcommand{\uhuru}{Uhuru}
\newcommand{\arielv}{Ariel-V}
\newcommand{\exosat}{EXOSAT}
\newcommand{\xrism}{XRISM}

\usepackage{subfigure}
\usepackage{amsmath}
\usepackage{xcolor}

\usepackage{graphicx}
\usepackage{wasysym}
\usepackage{txfonts}
\usepackage{xspace}
\usepackage{textcomp}
\usepackage{multirow}

\def\oldbf{}

\def\arraystretch{1.1}

\begin{document}
\title{%
  XMM-Newton Observations of Flares and a Possible Pulse Dropout in the Supergiant X-ray Binary 4U 1909+07}%

\author{Joel~B.~Coley\altaffilmark{1,2},
Ralf~Ballhausen\altaffilmark{3,4},
McKinley~Brumback\altaffilmark{5},  
Robin~H.D.~Corbet\altaffilmark{6,7},
Camille~M.~Diez\altaffilmark{8},
Felix~F\"urst\altaffilmark{8},
Nazma~Islam\altaffilmark{6},
Gaurava~K.~Jaisawal\altaffilmark{9},
Peter~Kretschmar\altaffilmark{8},
Christian~Malacaria\altaffilmark{10,11},
Katja~Pottschmidt\altaffilmark{2,6,$\dag$},
Pragati~Pradhan\altaffilmark{12}}
\email{joel.coley@howard.edu}
\altaffiltext{1}{Department of Physics and Astronomy, Howard University, Washington, DC 20059, USA}
\altaffiltext{2}{CRESST/Mail Code 661, Astroparticle Physics Laboratory, NASA Goddard Space Flight Center, Greenbelt, MD 20771, USA}
\altaffiltext{3}{Department of Astronomy, University of Maryland, College Park, MD 20742, USA}
\altaffiltext{4}{NASA-GSFC/CRESST, Astrophysics Science Division, Greenbelt, MD 20771, USA}
\altaffiltext{5}{Department of Physics, Middlebury College, 152 College Street Middlebury, Vermont 05753}
\altaffiltext{6}{University of Maryland Baltimore County, and X-ray Astrophysics Laboratory, Code 662 NASA Goddard Space Flight Center, Greenbelt Rd., MD 20771, USA}
\altaffiltext{7}{Maryland Institute College of Art, 1300 W Mt Royal Ave, Baltimore, MD 21217, USA}
\altaffiltext{8}{European Space Agency (ESA), European Space Astronomy Centre (ESAC), Camino Bajo del Castillo s/n, 28692 Villanueva de la Cañada, Madrid, Spain}
\altaffiltext{9}{{\oldbf DTU Space, Technical University of Denmark, \text{\O}rsteds Plads 348, DK-2800 Lyngby, Denmark}}
\altaffiltext{10}{International Space Science Institute, Hallerstrasse 6, 3012, Bern, Switzerland}
\altaffiltext{11}{{\oldbf INAF Osservatorio Astronomico di Roma, Via Frascati 33, 00078 Monte Porzio Catone (RM), Italy}}
\altaffiltext{12}{Department of Physics and Astronomy, Embry-Riddle Aeronautical University, 3700 Willow Creek Road, Prescott, AZ 86301, USA}
\altaffiltext{$\dag$}{deceased 17 June 2025}
\begin{abstract}

We report on a pair of \newton\ observations of the Supergiant X-ray binary 4U 1909$+$07, which were performed on 2021 October 3 and 2021 October 8, respectively.  We measure the neutron star rotation period in {\oldbf each observation to be $\sim$602.62\,s.}  This continues a long spin-up trend that has persisted since 2001 where the neutron star spin period was found to be {\oldbf $\sim$604.66\,s}.  In our timing analysis, we observe strong variations in the amplitude of the 1--10\,keV pulse profile as a function of time, and for the first time we find a low flux interval extending for a single pulse period in which pulsations are no longer detected.  We interpret this low flux interval as a pulse dropout similar to those observed in Vela X-1 and GX 301-2{\oldbf , which were each explained by a low-density cavity in the wind driving the propeller effect.}  In our time-resolved spectral analysis, we observed the spectral continuum, which can be described as an absorbed power law modified by a high-energy cutoff, to significantly soften during the pulse-dropout phase.  No evidence of an increasing absorption column density was found.  The observed softening {\oldbf in 4U 1909$+$07} {\oldbf also} supports an interpretation that the observed pulse dropout may be driven by the propeller effect{\oldbf , but the quasi-spherical settling accretion regime cannot be ruled out.}

\end{abstract}

\section{Introduction}
\label{Introduction}
Wind-fed Supergiant X-ray binaries (SGXBs) are comprised of a massive OB supergiant star, and in most cases, a highly magnetized neutron star orbiting their center of mass.  The neutron star accretes matter from the stellar wind of the OB supergiant companion, which has been observed to reach terminal velocities of several $\sim$1000\,km s$^{-1}$ and is often perturbed by inhomogeneous higher density and velocity structures referred to as clumps \citep{2008A&ARv..16..209P,2017SSRv..212...59M}.  In addition to clumps, optically thick, spiral shape corotating interaction regions (CIRs) extending up to tens of stellar radii may be ubiquitous in the wind of OB supergiant stars \citep{2014MNRAS.441.2173M}.  Due to the strong magnetic field of the neutron star, the accreted material is collimated to the poles of the neutron star and in the case that the magnetic and spin axes are misaligned, pulsed X-ray emission is observed.  The shape and strength of the pulsations are driven by the viewing geometry of the pulsar, its magnetic field geometry and donor star mass-loss rate, which for OB supergiant stars can reach up to 10$^{-4}$\,$M_\odot$ yr$^{-1}$ \citep[see e.g.][for a review]{2017SSRv..212...59M}.  While the shape and strength of pulsations are generally stable on short timescales, pulse-to-pulse variability is common in several X-ray binary pulsars \citep[e.g. Vela X-1; 1A 0535$+$262,][]{2014EPJWC..6406012K,2017A&A...608A.105B}.

In addition to pulse-to-pulse variability, pulse dropouts have been observed in several accreting X-ray pulsars where coherent pulsations are no longer detected on short timescales.  In highly absorbed environments such as an accretion disk or a dense wind, the non-detection of pulsations has been linked to obscuration effects where the pulsar beam is scattered out of the line of sight by the absorbed material.  This has been observed in the X-ray binary pulsars Her X-1 and SMC X-1, where the neutron star is periodically obscured by an edge-on warped precessing accretion disk resulting in superorbital modulation on timescales of tens of days \citep{2002astro.ph..3213K,2019ApJ...875..144P}, as well as wind-fed SGXBs due to clumpy material \citep[e.g. OAO 1657-415;][]{2014MNRAS.442.2691P}.  In other cases, pulse dropout may be linked to a large reduction in the mass accretion rate on two orders of magnitude or more \citep[see e.g. V 0332$+$53, 4U 0115$+$63;][]{2016A&A...593A..16T}.  A mechanism that may explain pulse dropouts that are linked to a variable accretion rate include the ``propeller effect'' where the ram pressure driven by accretion no longer overcomes the pressure caused by the magnetic field of the neutron star  \citep{1975A&A....39..185I}.  In some cases the pulse dropout may possibly be driven by low-density cavities in a clumpy wind environment \citep[e.g. Vela X-1;][]{2008A&A...492..511K,2015A&A...575A..58M} or is accompanied by the quasi-spherical subsonic accretion regime where the accreted material no longer penetrates the magnetosphere of the neutron star due to inefficient cooling likely due to Compton processes \citep{2012MNRAS.420..216S}.


\begin{table}
  \caption{Relevant orbital parameters {\oldbf of 4U 1909$+$07}}
  \label{tab:orb}

  \centering
  \begin{tabular}{ccccc}
                \hline\hline
                      Parameter & Orbital second-epoch & Units & References \rule[0mm]{0mm}{3mm}\\
                       & Values &  &  \rule[0mm]{0mm}{3mm}\\
                \hline
    $a_x$ $\sin i$    & 47.83\,$\pm$\,0.94 & lt-sec & \citet{2004ApJ...617.1284L} \rule[0mm]{0mm}{3mm}\\
    $\tau_{90}$     & 52,631.383  $\pm$ 0.013 & MJD & \citet{2004ApJ...617.1284L} \\
    $P_{\rm spin}$ & 604.684  $\pm$ 0.001 & seconds & \citet{2004ApJ...617.1284L} \\
    $\dot{P_{\rm spin}}$ & (1.22 $\pm$ 0.09) $\times$ 10$^{-8}$ & \rm{s} \rm{s}$^{-1}$ & \citet{2004ApJ...617.1284L} \\
    $e$ & 0.021 $\pm$ 0.039 & & \citet{2004ApJ...617.1284L} \\ 
    $P_{\rm orb}$      & 4.4007 $\pm$ 0.0009 & days & \citet{2004ApJ...617.1284L} \\
                $f(M)$      & 6.07 $\pm$ 0.35 & $M_{\odot}$ & \citet{2004ApJ...617.1284L} \\
                \hline
    $P_{\rm super}$      & 15.196\,$\pm$\,0.004 & days & \citet{2023ApJ...948...45I} \\                  
                \hline
    $i$      & 46--58 & degrees & \\
    Spectral Type & B0-B3 & & \\            
                \hline
  \end{tabular}

\end{table}

4U 1909$+$07 is a SGXB consisting of a neutron star rotating at a period of $\sim$604\,s \citep{2004ApJ...617.1284L} and a massive early-type star orbiting their center of mass with a period of $\sim$4.4\,days \citep{2000ApJ...532.1119W,2013ApJ...778...45C}.  {\oldbf It additionally shows superorbital modulation on a timescale of 15.196\,$\pm$\,0.004\,days, as was observed by the \swift\ BAT \citep{2013ATel.5119....1C,2023ApJ...948...45I,2025arXiv250504452R}.}  4U 1909$+$07 was first discovered by the \uhuru\ satellite in 1974 \citep{1974ApJS...27...37G}, and was subsequently detected by the MIT/OSO 7 \citep{1979ApJS...39..573M}, \arielv\ \citep{1984ApJS...56..507W}, HEAO 1 \citep{1984ApJS...56..507W}, and \exosat\ \citep{1988MNRAS.232..551W} satellites.  Using $H$- and $K$- band spectroscopy from the Long-slit Intermediate Resolution Infrared Spectrograph (LIRIS) mounted on the 4.2\,m William Herschel Telescope (WHT), the spectral type of the donor star was found to be in the range of a B0 I to B3 I \citep{2015A&A...578A.107M}.  The distance to the source was estimated to be 4.85\,$\pm$\,0.50\,kpc, which is consistent with recent parallax measurements {\oldbf corresponding to a distance of 5.0$^{+3.9}_{-2.1}$\,kpc} reported in the GAIA DR3 catalog \citep{2021AJ....161..147B,2023A&A...674A...1G}.  {\oldbf Due to the smaller uncertainty reported in the WHT estimate, we adopt the distance estimate reported in \citet{2015A&A...578A.107M} in this work.}  Its orbital ephemeris derived from a pulsar timing analysis from \citet{2004ApJ...617.1284L} is reported in Table~\ref{tab:orb}.

A soft X-ray spectral analysis using the \chandra\ High Energy Transmission Gratings \citep[HETG;][]{2005PASP..117.1144C} revealed strong Fe K$\alpha$ and Fe K$\beta$ lines {\oldbf at $\sim$6.4\,keV and $\sim$7.05\,keV, respectively,} along with an Fe K edge at $\sim$7.1\,keV \citep{2010ApJ...715..947T}.  They found the Fe K$\beta$ to Fe K$\alpha$ line flux ratio to be $\sim$0.3, which is significantly larger than the theoretical value of $\sim$0.13--0.14 reported in \citet{2003A&A...410..359P} for neutral to moderately ionized plasmas.  The Fe K$\alpha$ and Fe K$\beta$ lines were also observed using \suzaku\ \citep{2012A&A...547A...2F,2013ApJ...779...54J}, \astrosat\ \citep{2020MNRAS.498.4830J}, and \nustar\ \citep{2020MNRAS.498.4830J,2023ApJ...948...45I}.

In addition to Fe K$\alpha$ and Fe K$\beta$ lines, the \chandra\ observations showed evidence of a Compton shoulder on the soft energy edge of the Fe K$\alpha$ line \citep{2010ApJ...715..947T}, which is consistent with 4U 1909$+$07 being embedded in a Compton-thick medium.  The flux ratio of the Compton shoulder relative to the primary Fe K$\alpha$ line was found to be $\sim$0.48.  It was detected at {\oldbf a neutral hydrogen column density, $N_{\rm H}$,} of $\sim$4$\times$10$^{22}$\,cm$^{-2}$, but was not found in two \chandra\ observations where $N_{\rm H}$ was measured to be $\sim$1.5$\times$10$^{23}$\,cm$^{-2}$ and $\sim$2.9$\times$10$^{23}$\,cm$^{-2}$, respectively \citep{2010ApJ...715..947T}.  The non-detection of the Compton shoulder at higher column densities may be linked to smearing effects due to second-order scatterings between the fluorescence photons and electrons in the ambient medium \citep{2002MNRAS.337..147M}.

Since the discovery of its $\sim$604\,s spin period, the neutron star in 4U 1909$+$07 was found to show long-term changes in its rotation period.  Timing measurements using \rxte, \integral\ and \suzaku\ observations showed that the neutron star rotation period changes erratically on a timescale of years \citep{2011A&A...525A..73F,2012A&A...547A...2F,2013ApJ...779...54J}, which shows a trend that is consistent with a random walk \citep{1993MNRAS.262..726D}.  Using \nustar\ and \astrosat\ observations performed on MJD\,57,204.65 and MJD\,57,951.27, respectively, \citet{2020MNRAS.498.4830J} extended this baseline and found the average spin-up rate $\dot{P}$ between 2001 and 2017 to be 1.71$\times$10$^{-9}$ s s$^{-1}$.  Recently, \citet{2023ApJ...948...45I} refined the average spin-up rate $\dot{P}$ between 2001 and 2019 to be 4.7$\times$10$^{-9}$\,s\,s$^{-1}$ using two additional \nustar\ observations.  This indicates that while the torque of the neutron star shows erratic behavior on a timescale of years, it exhibits clear spin-up on the longest timescales.

In this paper, we present timing and spectral analyses of two \newton\ observations of 4U 1909$+$07.  The remainder of the paper is organized as follows. The \xmm\ observations are presented in Section~\ref{Data Analysis and Modeling}.  In Section~\ref{Timing Analysis}, we measure the neutron star rotation period and for the first time we observe the evidence of a dropout of pulsations spanning a single $\sim$602\,s spin cycle.  The time-averaged and time-resolved spectral analysis is presented in Sections~\ref{Averaged Spectral Analysis}--~\ref{Pulse-to-Pulse Spectral Analysis}.  We provide a discussion of the results in Section~\ref{Discussion} and the conclusions are given in Section~\ref{Conclusion}.  If not stated otherwise, the uncertainties and limits presented in the paper are at the 90$\%$ confidence level.

\section{Data and Analysis}
\label{Data Analysis and Modeling}

\begin{table}
  \caption{Summary of \xmm\ Observations of 4U 1909$+$07}
  \label{X-ray Observations Summary}

  \centering
  \begin{tabular}{ccccc}
    \hline\hline
    ObsID & Start Time & End Time & Orbital & Exposure \\
     & (MJD) & (MJD) & Phase$^a$ & (ks) \\
    \hline
    0884850201 & 59490.594 & 59491.109 & 0.664--0.781 & 46.9 \\
    0884850301 & 59495.205 & 59495.457 & 0.712--0.769 & 21.6 \\
    \hline
  \end{tabular}
  \tablecomments{\\*
    $^a$ Orbital phase zero is defined at MJD\,52,631.383, corresponding to the time of mean longitude, $T_{\pi/2}$ \citep{2004ApJ...617.1284L}.}

\end{table}

\begin{figure}[h]
\centerline{\includegraphics[width=3.3in]{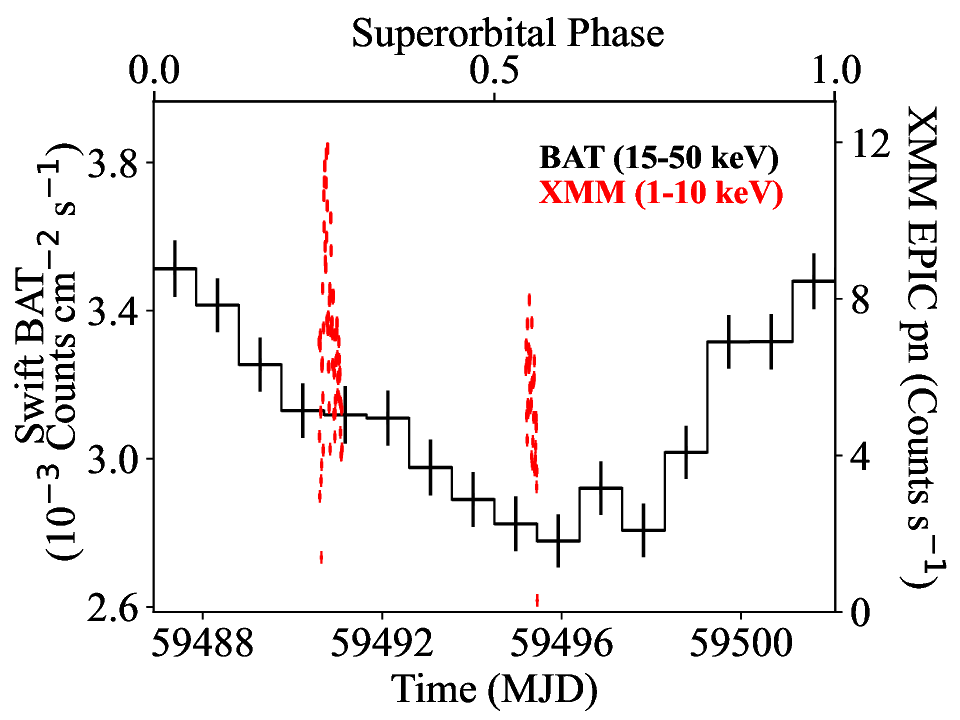}}
\caption[f1.eps]{
\swift\ BAT light curve (black) folded on the $\sim$15.2\,day superorbital period using 16 bins.  Phase zero corresponds to the time of maximum flux \citep{2023ApJ...948...45I}.  The \xmm\ EPIC-pn 1--10\,keV light curves for observations I and II binned to a time resolution of {\oldbf 602.62\,s} are overplotted in red (see text for details).}
\label{4U 1909 Figure 1}
\end{figure}

Our \xmm\ campaign consisted of two observations that were performed at {\oldbf orbital phases 0.664--0.781 during two consecutive orbital cycles,} but sampled different phases of a single $\sim$15.2\,day superorbital cycle.  An observation log is given in Table~\ref{X-ray Observations Summary}.  In {\oldbf Figure~\ref{4U 1909 Figure 1}}, we show the \swift\ BAT light curve {\oldbf folded on the $\sim$15.2\,day superorbital period}.  The first observation (hereafter Observation I) was performed on 2021 October 3--4 during the decay transition between the maximum and minimum of the $\sim$15.2\,day period, and the second observation (hereafter Observation II) took place on 2021 October 8 during the superorbital minimum.

\xmm\ \citep{2001A&A...365L...1J} carries two X-ray instruments: the European Photon Imaging Camera, which consists of two MOS cameras \citep[EPIC-MOS;][]{2001A&A...365L..27T} and one pn camera \citep[EPIC-pn;][]{2001A&A...365L..18S}, and the Reflection Grating Spectrometer \citep[RGS;][]{2001A&A...365L...7D}.  Due to the large intrinsic absorption of the source on the order of 10$^{23}$\,cm$^{-2}$ \citep[see e.g.][]{2010ApJ...715..947T,2012A&A...547A...2F}, the RGS data were not usable and will not be discussed further in the paper.  We reduced and screened the \xmm\ data using the Science Analysis System (SAS) version 20.0.0 following standard reprocessing and screening criteria.

To mitigate effects due to pileup, the EPIC-pn and EPIC-MOS 2 instruments were performed in timing mode with a time resolution of 30\,$\mu$s and 1.75\,ms, respectively.  Due to {\oldbf dead chips} resulting from micrometeorite impacts in 2005 March and 2012 December\footnote{https://heasarc.gsfc.nasa.gov/docs/xmm/uhb/epic.html}, the EPIC-MOS 1 instrument was operated in Small Window mode with a time resolution of 0.3\,s.  For both observations, the EPIC-pn and EPIC-MOS 1 cameras adopted the thick filters, while EPIC-MOS 2 used the medium filter.

For EPIC-MOS 1, we initially extracted the source spectra from a circular region of radius 30$\arcsec$ centered on the source, adopting a PATTERN selection from 0 to 12.  The background spectra {\oldbf were extracted} from a circular region of radius 30$\arcsec$ offset from the source. {\oldbf To investigate if the spectral parameters were dependent on the choice of background, we also tested background regions of radii 45$\arcsec$ and 60$\arcsec$ offset from the source.  We found the spectral parameters do not significantly depend on the choice of the background (see Sections~\ref{Averaged Spectral Analysis}--\ref{Pulse-to-Pulse Spectral Analysis}).}  We checked for pileup using the XMMSAS tool \texttt{epatplot} and found significant pileup, suggesting that we should excise the core in the point-spread function (PSF).  To significantly reduce the presence of pileup, we extracted the source spectra from an annular region with an inner radius of 8$\arcsec$ and an outer radius of 30$\arcsec$ and only used a PATTERN selection of 0 to contain single events.  We produced light curves in the 10--12\,keV band and found no evidence of soft proton background flares.

Since EPIC-MOS 2 and EPIC-pn were both operated in timing mode, we similarly checked for pileup using \texttt{epatplot} and found none, which is not surprising given the {\oldbf moderate luminosity of the source.}  For EPIC-pn, the source spectra were extracted using Columns 30--44 and were filtered using a PATTERN selection from 0 to 4 to contain only single and double events.  The background spectra were extracted using EPIC-pn Columns 3--13.  For EPIC-MOS 2, the source spectra were extracted using Columns 290--320 and were filtered using a PATTERN selection from 0 to 12.  The background was extracted from a circular region of radius 30$\arcsec$ offset from the source.  We similarly checked for soft proton background flares and found none.

We generated response matrices using the packages \texttt{RMFGEN} and \texttt{ARFGEN}.  For the time-resolved spectral analysis (see Section~\ref{sec:ks-integrated}), we generated Good Time Intervals (GTIs) using the SAS task \texttt{gtibuild}.

Event times were corrected to the solar system barycenter using the tool XMMSAS tool \texttt{barycen} with the DE-200 solar system ephemeris.  We further corrected the event times for the orbital motion of the neutron star using the ephemeris defined in \citet{2004ApJ...617.1284L}.  We used the \texttt{EPICLCCORR} task to correct for remaining instrumental effects (e.g. vignetting).

The time-averaged net count rates from the source over the full energy range during Observations I were found to be as follows: 6.69\,$\pm$\,0.01\,counts s$^{-1}$ (EPIC-pn), 0.488\,$\pm$\,0.003\,counts s$^{-1}$ (EPIC-MOS 1, PSF excised) and 1.590\,$\pm$\,0.007\,counts s$^{-1}$ (EPIC-MOS 2), respectively.  For Observation II, the time-averaged net count rates were found to be as follows: 5.24\,$\pm$\,0.02\,counts s$^{-1}$ (EPIC-pn), 0.372\,$\pm$\,0.004\,counts s$^{-1}$ (EPIC-MOS 1, PSF excised), and 1.186\,$\pm$\,0.007\,counts s$^{-1}$ (EPIC-MOS 2), respectively.  For the remainder of the paper, we rebinned the spectral file using the optimal binning regime using the FTOOL \texttt{ftgrouppha} as described in \citet{2016A&A...587A.151K} if not stated otherwise.


\section{Timing Analysis}
\label{Timing Analysis}

\subsection{Light Curve and Pulse Profiles}
\label{Temporal Analysis}

To determine the neutron star rotation period in the \xmm\ observations, we used the epoch folding technique presented in \citet{1987A&A...180..275L}, {\oldbf which we} applied to the EPIC-pn light curves, binned to a resolution of 1\,s after correcting for the binary orbital motion.  We estimated the uncertainty on the pulse period at the 1$\sigma$ confidence interval by simulating 2000 light curves based on the previously determined pulse period and profile with additional Poisson noise.  We find the neutron star rotation period during observations I and II to be 602.62\,$\pm$\,0.02\,s and 602.62\,$\pm$\,0.07\,s, which {\oldbf is unchanged within the uncertainties.}

\begin{figure}[h]
\centerline{\includegraphics[width=3in]{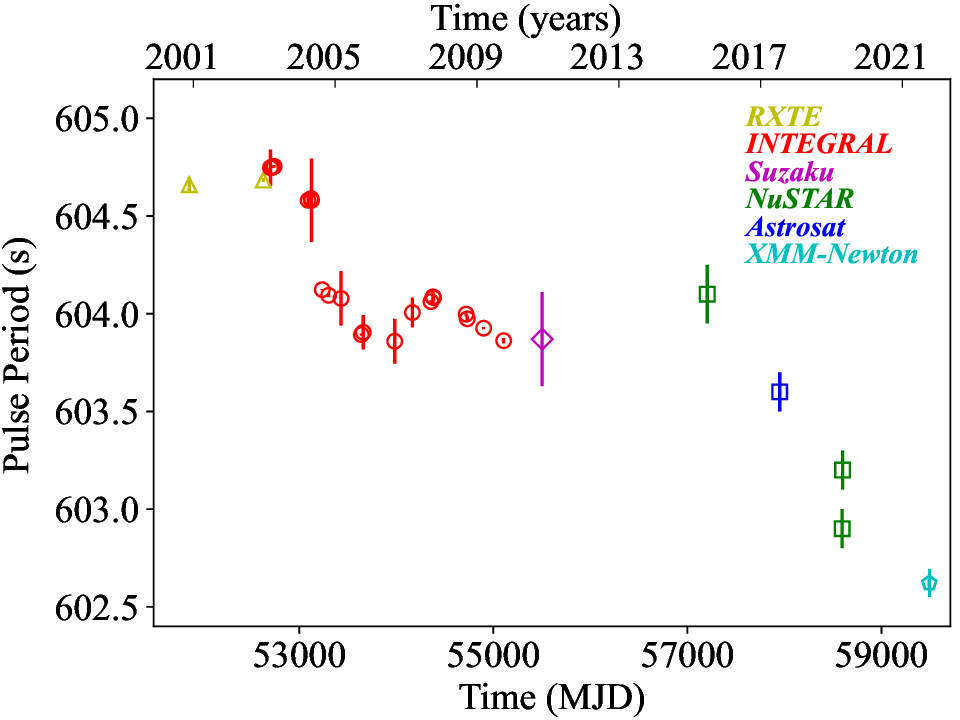}}
\caption[f2.eps]{
Long-term evolution of the neutron star rotation period between {\oldbf 2001}--2021.  The pulse period measured by \rxte\ \citep{2004ApJ...617.1284L}, \integral\ \citep{2011A&A...525A..73F}, \suzaku\ \citep{2012A&A...547A...2F,2013ApJ...779...54J}, \nustar\ \citep{2020MNRAS.498.4830J,2023ApJ...948...45I}, \astrosat\ \citep{2020MNRAS.498.4830J} and \xmm\ (this work) are indicated by the gold triangles, red circles, magneta diamond, green squares, blue squares and cyan plus, respectively.
\label{4U 1909 Figure 2}
}
\end{figure}

\begin{figure*}
  \centering
  \includegraphics[width=.45\linewidth]{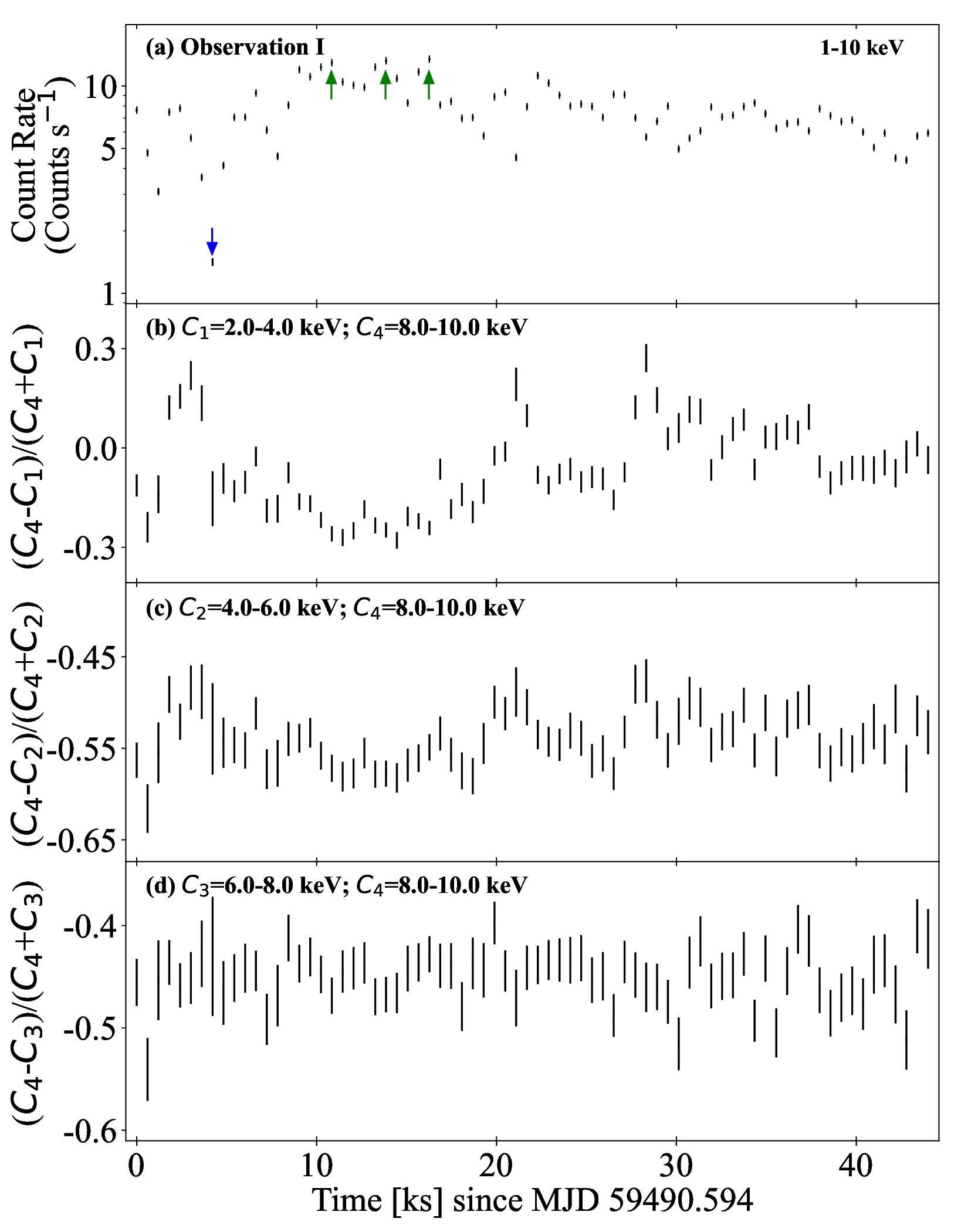} \qquad
  \includegraphics[width=.45\linewidth]{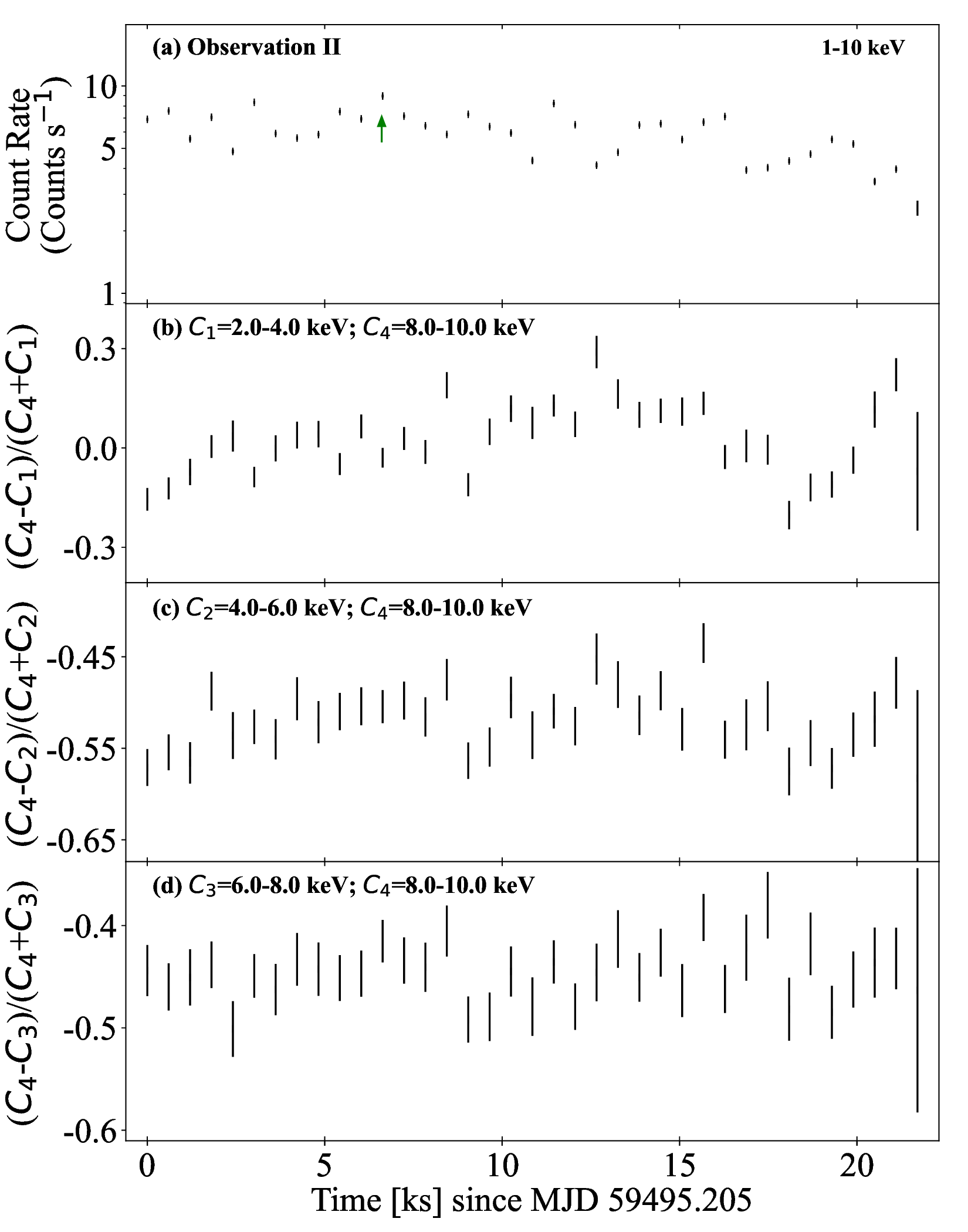}
  \caption{(a) \xmm\ EPIC-pn 1--10\,keV light curves of 4U 1909$+$07 for Observations I (left) and II (right).  {\oldbf The time of the off-state is indicated by the blue arrow.  Short flares are indicated by green arrows.}  The time resolution is the local pulse period (see text for details).  Count rate is plotted on a logarithmic scale to emphasize the energy-dependence of the variability.  The hardness ratios of the 2--4\,keV and 8--10\,keV, 4--6\,keV and 8--10\,keV, and 6--8\,keV and 8--10\,keV bands are shown in panels {\oldbf (b), (c) and (d)}, respectively.}
    \label{4U 1909 Figure 3}
\end{figure*}

Figure~\ref{4U 1909 Figure 2} shows the long-term evolution of the pulse period between 2001 and 2021, which extends the pulse period history presented in \citet{2023ApJ...948...45I} by an additional $\sim$895\,days.  We find that the neutron star rotation period changed from 604.66\,$\pm$\,0.04\,s to 602.62\,$\pm$\,0.07\,s between 2001 and 2021.  To investigate changes in the neutron star rotation period between our \xmm\ observations and the spin period reported using \rxte\ in \citet{2004ApJ...617.1284L}, we calculated the Taylor expansion,

\begin{equation}
\label{Period Derivative Calculation}
P(t)=P(t_0)+(t-t_0) \dot{P}
\end{equation}
where $P(t)$ is the neutron star rotation period derived during Observation II, $P(t_0)$ is the neutron star rotation period derived in \citet{2004ApJ...617.1284L}, and the epoch $t_0$ is the maximum delay time from the pulsar timing analysis {\oldbf (see Table~\ref{tab:orb})}.  We find the pulse period derivative between the \rxte\ observations and our \xmm\ observation to be (-3.09\,$\pm$\,0.07)$\times$10$^{-9}$\,s\,s$^{-1}$.

\begin{figure}[h]
\centerline{\includegraphics[width=3in]{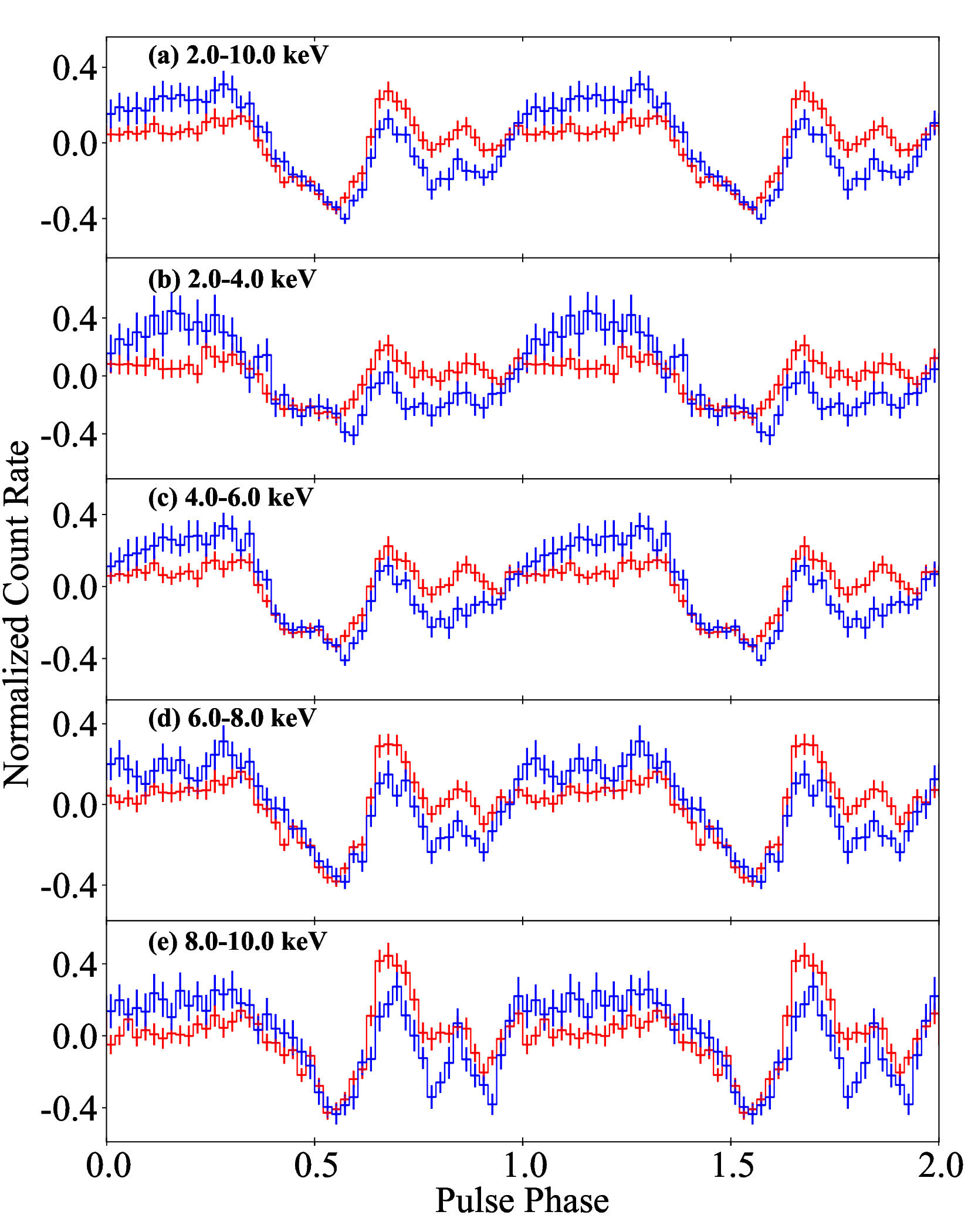}}
\caption[f4.eps]{
Energy-resolved \xmm\ EPIC-pn pulse profiles for Observations I (red) and II (blue).  The pulse profiles are normalized such that their mean value is zero and their standard deviation is unity.  The modulation appears to show a broad plateau and a narrow dip feature.
\label{4U 1909 Figure 4}
}
\end{figure}

\begin{figure*}[ht]
\centering
\begin{tabular}{cc}
	\includegraphics[trim=0cm 0cm 0cm 0cm, clip=false, scale=0.4, angle=0]{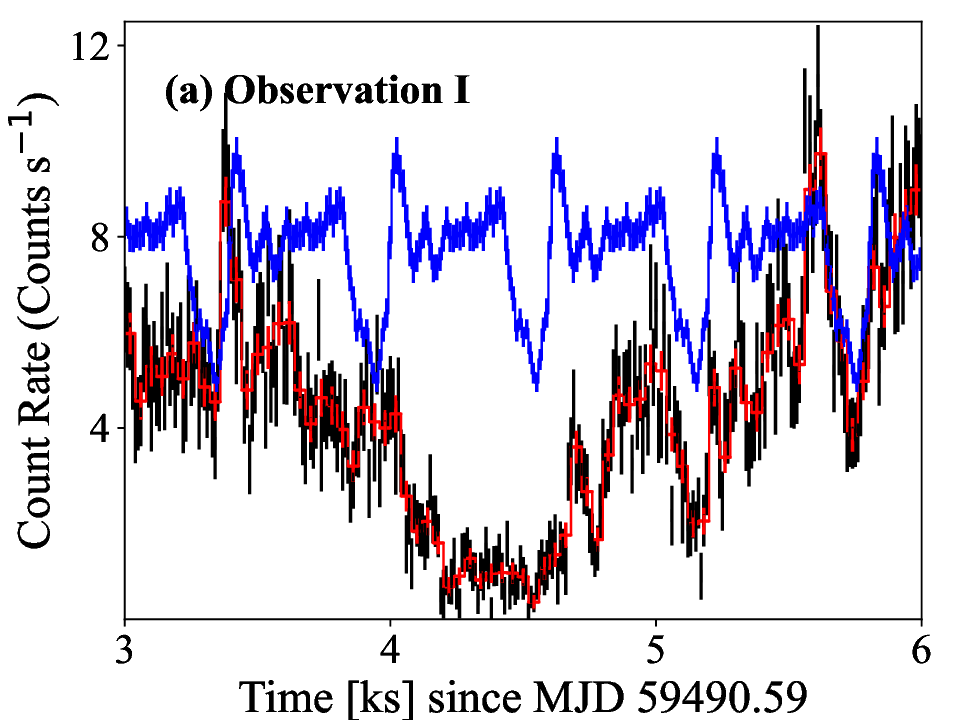} & \includegraphics[trim=0cm 0cm 0cm 0cm, clip=false, scale=0.4, angle=0]{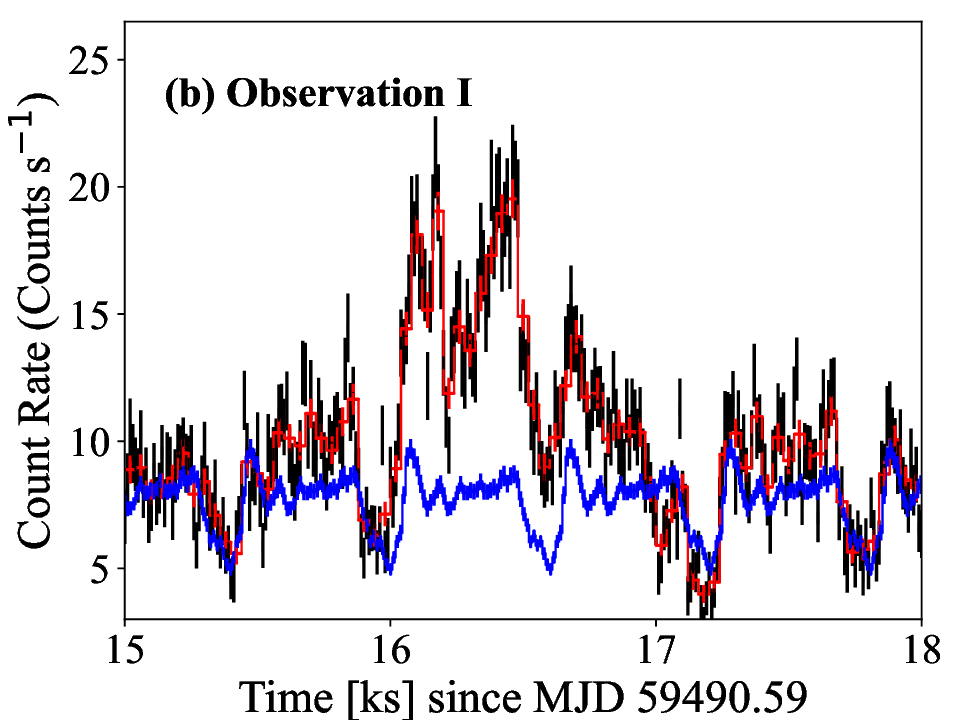} \\
	\includegraphics[trim=0cm 0cm 0cm 0cm, clip=false, scale=0.4, angle=0]{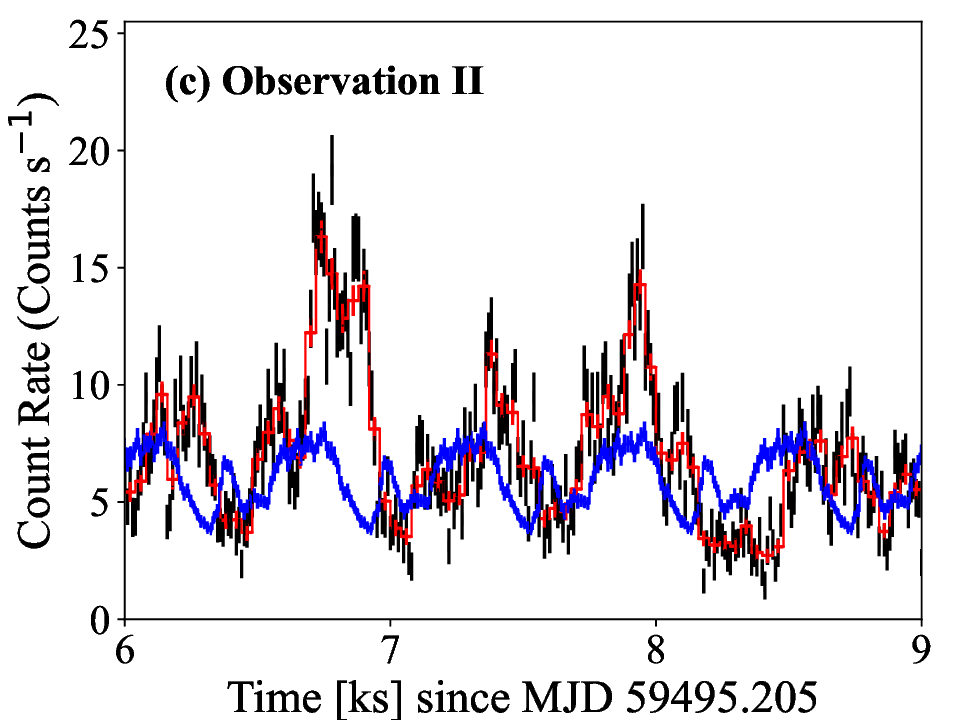} & \includegraphics[trim=0cm 0cm 0cm 0cm, clip=false, scale=0.4, angle=0]{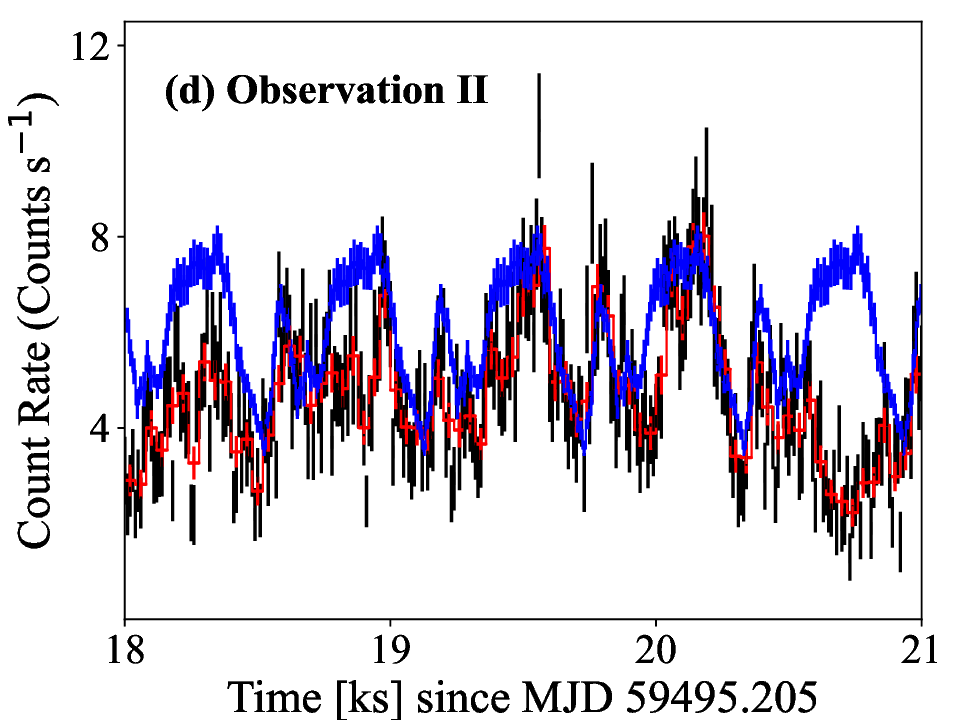}
\end{tabular}
\caption{3\,ks segments of the 1--10\,keV EPIC-pn light curve {\oldbf for Observations I (top) and II (bottom).  Panels (a) and (b) show the off state and a flare, respectively.  Panels (c) and (d) show a flare and a drop in the count rate that is not coincident with a pulse dropout, respectively.}  The light curves binned at a resolution of 10\,s and 40\,s are indicated in black and red, respectively.  The average pulse profile for the full observation is overplotted in blue.  Note the $y-$axes are not the same scale.}
\label{4U 1909 Figure 5}
\end{figure*}

\begin{figure*}[ht]
    \centering
    \begin{tabular}{c}
    \subfigure
    {
        \includegraphics[width=0.6\textwidth]{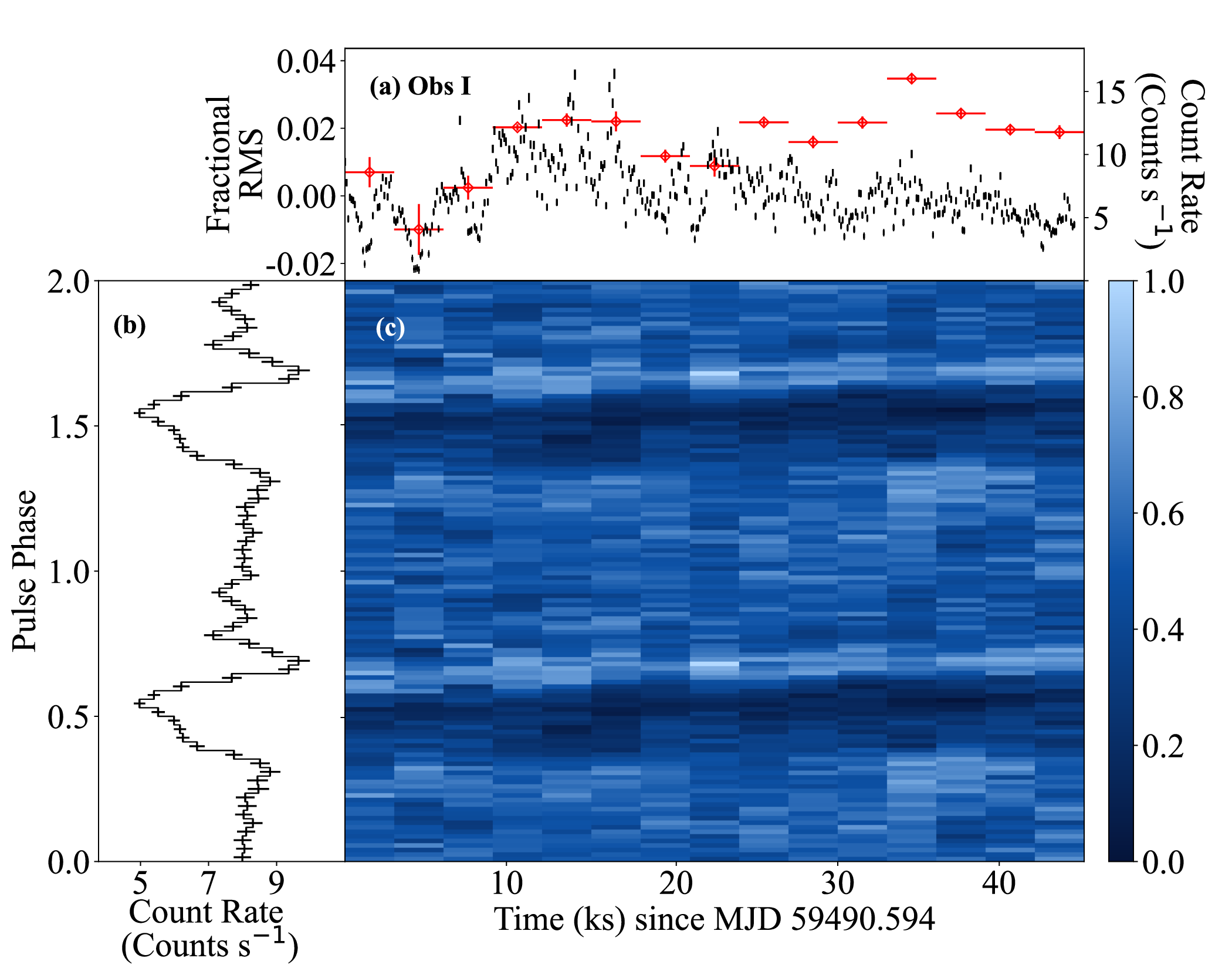}
        \label{Obs I Time Evolution}
    } \\
    {
        \includegraphics[width=0.6\textwidth]{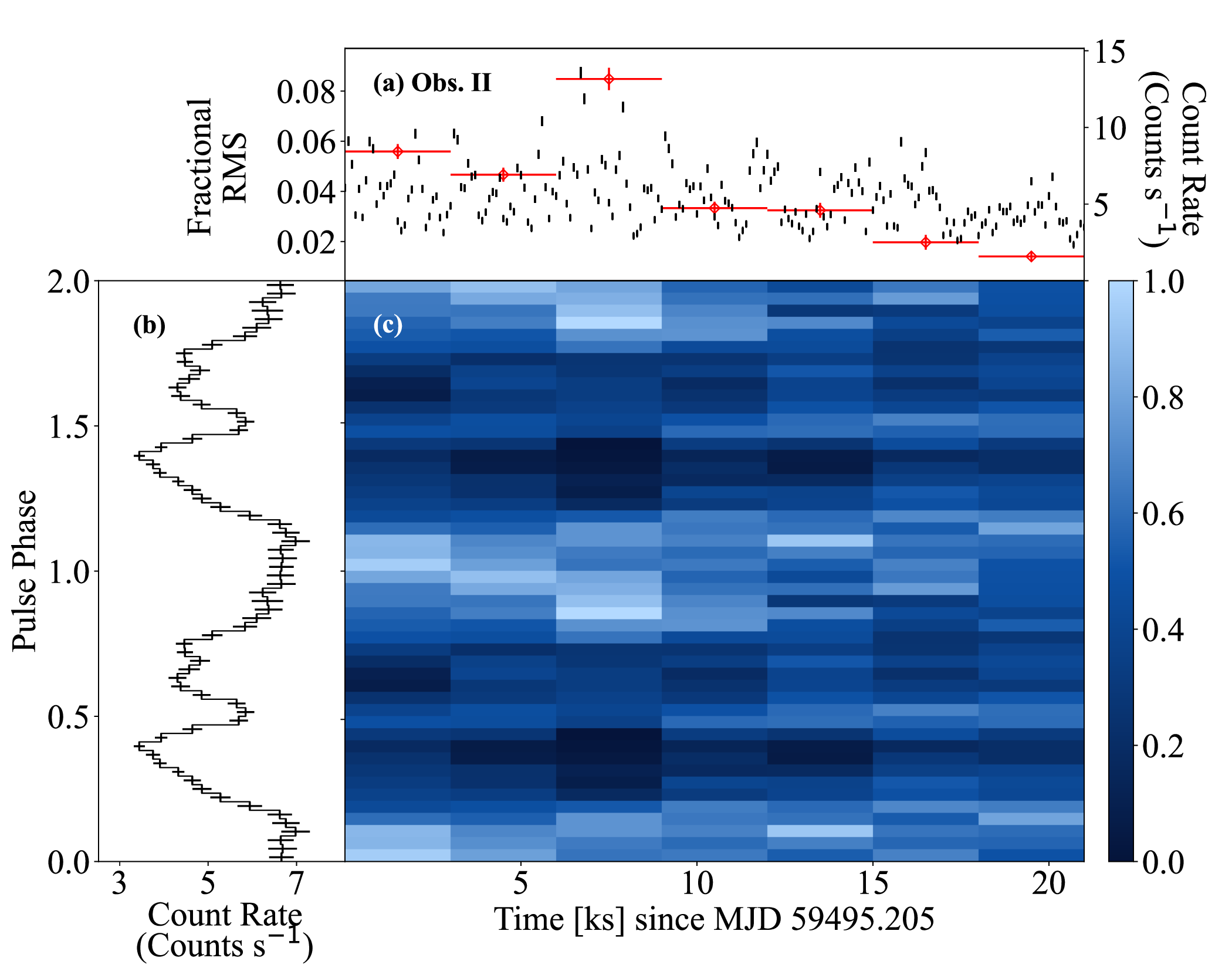}
        \label{ObsII Time Evolution}
    }
    \end{tabular}
    \caption{(a) Variation in the fractional root mean square (rms) amplitude for Observation I (top) and Observation II (bottom) of the $\sim$602\,s pulsation cycle as a function of time (see text for details).  The fractional rms{\oldbf , indicated in red,} was calculated for 3\,ks time intervals.  The \xmm\ EPIC-pn {\oldbf count rates} are overplotted (right axis). (b) EPIC-pn pulse profile for the full observation folded on the $\sim$602\,s neutron star rotation period. (c) Pulse profile map scaled from zero to one, where the pulse profiles are calculated for 3\,ks time intervals (see text for {\oldbf details)}.}
    \label{4U 1909 Figure 6}
\end{figure*}

In Figure~\ref{4U 1909 Figure 3}(a), we plot the 1--10\,keV EPIC-pn light curves of 4U 1909$+$07 for Observations I and II binned at a time resolution chosen to be the local pulse period.  These revealed large variations in flux on kilosecond timescales, similar to that previously observed by \suzaku\ \citep{2012A&A...547A...2F}, \nustar\ and \astrosat\ \citep{2020MNRAS.498.4830J,2023ApJ...948...45I} as well as other classical SGXBs \citep[e.g. OAO 1657-415; Vela X-1,][]{2023ApJ...945...51P,2023A&A...674A.147D}.  In Observation I, we observe {\oldbf three flares, which we indicate by green arrows.  The times of the flares are $T_{\rm obs}\approx$10.85\,ks into the observation, $T_{\rm obs}\approx$13.86\,ks into the observation and $T_{\rm obs}\approx$16.27\,ks into the observation.  The 1--10\,keV count rates for the flares were found to be 13.0\,$\pm$\,0.2\,counts s$^{-1}$, 13.2\,$\pm$\,0.2\,counts s$^{-1}$ and 13.5\,$\pm$\,0.2\,counts s$^{-1}$. Additionally, we observe a short dip $\sim$4.22\,ks into the observation, which we indicate by the blue arrow.  The observed 1--10\,keV count rate in the dip drops to 1.42\,$\pm$\,0.06\,counts s$^{-1}$.}  Unless otherwise stated, we refer to the dip as an off state.  In Observation II, we find {\oldbf one moderate flare about $\sim$6.63\,ks into the observation reaching a peak count rate of 8.9\,$\pm$\,0.1\,counts s$^{-1}$}, but no dips comparable to those found in the first observation were detected.

{\oldbf We folded the 2--10\,keV light curves for Observations I and II on the 602.62\,$\pm$\,0.02\,s and 602.62\,$\pm$\,0.07\,s periods, respectively. For Observation I, we defined phase zero at MJD\,52,631.383, corresponding to the time of mean longitude.  We aligned the pulse profiles for both observations by calculating the maximum value of the cross correlation function between the two pulse profiles (see Figure~\ref{4U 1909 Figure 4}(a)).  The pulse profiles for each observation show several dips and plateaus.

In Figure~\ref{4U 1909 Figure 5}(a), we zoom in on the dip in Observation I binned to a time resolution of 10\,s and 40\,s.  Significant variability is observed during the dip, including a $\sim$602\,s segment where the count rate drops to a minimum of 0.84\,$\pm$\,0.09\,counts s$^{-1}$ and where pulsations are not detected.  We interpret this as a possible pulse dropout, where no visible pulsations of the neutron star are observed.  Apart from the $\sim$602\,s interval, clear pulsations are observed at the start and end of the light curve segment.  We used epoch folding to calculate the pulse period from the beginning of the observation to $\sim$4.04\,ks and from $\sim$4.64\,ks to the end of the observation, resulting in an exposure of $\sim$39.8\,ks.  We find the pulse period in these segments to be 602.5\,$\pm$\,0.3\,s and 602.64\,$\pm$\,0.04\,s, which is consistent with our value of $\sim$602.62\,s for the full observation.
  
In Figures~\ref{4U 1909 Figure 5}(b) and~\ref{4U 1909 Figure 5}(c), we plot 10\,s and 40\,s light curves of the third flare in Observation I and the largest flare in Observation II, respectively.  We find significant variability superimposed on the flares.  For the third flare in Observation I, we observe the count rate increases to a maximum of 21.2\,$\pm$\,1.6\,counts s$^{-1}$ for about a $\sim$602\,s pulse period and decreases again to an average of $\sim$7.1\,counts s$^{-1}$ following the pulse.  Similarly, for the largest flare in Observation II the count rate reaches a maximum of 19.2\,$\pm$\,1.5\,counts s$^{-1}$ for about a $\sim$602\,s pulse period.  Finally, we show the EPIC-pn light curve between 18\,ks and 21\,ks in Figure~\ref{4U 1909 Figure 5}(d).  We find that while the count rate was significantly decreased, reaching a minimum of 1.3\,$\pm$\,0.5\,counts s$^{-1}$, no evidence of a pulse dropout was observed.}
  
We additionally divided the EPIC-pn light curves for both observations into four energy bins defined between energies of 2--4\,keV, 4--6\,keV, 6--8\,keV and 8--10\,keV to investigate the energy dependence of the source variability.  Unless otherwise stated, we denote the count rate in the 2--4\,keV, 4--6\,keV, 6--8\,keV and 8--10\,keV bands as $C_{1}$, $C_{2}$, $C_{3}$ and $C_{4}$.

In Figure~\ref{4U 1909 Figure 4}{\oldbf (b)--(c)--(d)--(e)}, we show the pulse profiles in the $C_{1}$, $C_{2}$, $C_{3}$ and $C_{4}$ bands.  While the pulse profiles only show marginal evidence of an energy dependence in the \xmm\ band, we note that at harder energies the pulse profile evolves from a broad plateau-like peak with several dips below $\sim$10\,keV to a single peaked profile above $\sim$20\,keV \citep{2012A&A...547A...2F,2013ApJ...779...54J}.

Comparing Observations I and II, we observe {\oldbf that the overall} shape of the pulse profile {\oldbf is} similar although {\oldbf the relative strength of the components between pulse phases 0.0--0.3 and 0.6--0.9 shows significant differences (see Figure~\ref{4U 1909 Figure 4}).}  This is similar to {\oldbf changes in the pulse profile morphology} that was reported in \citet{2020MNRAS.498.4830J}, using \nustar\ and \astrosat\ observations performed in 2015 and 2017, respectively.  We additionally compare the pulse profiles in each of our \xmm\ observations to those reported in archival \rxte\ \citep{2011A&A...525A..73F}, \suzaku\ \citep{2012A&A...547A...2F}, \astrosat\ \citep{2020MNRAS.498.4830J} and \nustar\ \citep{2020MNRAS.498.4830J,2023ApJ...948...45I} data.  We also find apart from {\oldbf differences in the relative strength of components at pulse phases 0.0--0.3 and 0.6--0.9}, the {\oldbf overall shape} of the 3--10\,keV pulse profiles is similar to archival observations.  This may suggest that while 4U 1909$+$07 shows a global spin-up trend, a similar emission geometry persisted on a timescale of years.

As indicated in Section~\ref{Introduction}, pulse-to-pulse variability is common in X-ray binary pulsars \citep[e.g. Vela X-1; 1A 0535$+$262,][]{2014EPJWC..6406012K,2017A&A...608A.105B}.  To monitor short-term changes in the shape and amplitude of the pulse profile, we constructed a pulse profile map for each observation using the EPIC-pn light curves (see Figure~\ref{4U 1909 Figure 6}(c)).  We derived pulse profiles for different segments of the light curve, and successively shift the start time of each segment relative to the previous one by an offset in time.  We experimented with different light curve segments and found 3\,ks for both the segment length{\oldbf ,} as well as the offset{\oldbf ,} is a good compromise to show changes in the strength and shape of the pulse profiles in both observations.  This resulted in 15 statistically independent light-curve segments for Observation I{\oldbf ,} and 7 statistically independent light-curve segments for Observation II.

For Observation I, we folded the light curve from each segment of data on the 602.62\,$\pm$\,0.02\,s neutron star rotation period determined from the full observation and find that the pulse profile shape dramatically changes across the observation (see Figure~\ref{4U 1909 Figure 6}, top panel).  We find that the pulse profile amplitude is initially weak in the first 9\,ks, and shows a dramatic increase coincident with an increase in the source flux.  In Observation II, we similarly fold the light curve from each segment of data on the 602.62\,$\pm$\,0.07\,s period determined from the full observation.  The shape of the pulse profiles show a similar, but less dramatic, evolution compared to Observation I (see Figure~\ref{4U 1909 Figure 6}, bottom panel).

{\oldbf \subsection{Hardness Ratios}
\label{Hardness Ratios}
Since flux variability is often tied with changes in the spectral shape, we plot the hardness ratios in the 2--4\,keV, 4--6\,keV{\oldbf ,} and 6--8\,keV bands relative to the 8--10\,keV band in {\oldbf Figure~\ref{4U 1909 Figure 3}(b)-(c)-(d)}.  We define the hardness ratios as:

\begin{equation}
\label{Hardness Equation}
{\rm HR}=\frac{(C_{\rm hard}-C_{\rm soft})}{(C_{\rm hard}+C_{\rm soft})}.
\end{equation}

We find strong variations in the hardness ratio between $C_{1}=$2--4\,keV and $C_{4}=$8--10\,keV bands{\oldbf ,} and $C_{2}=$4--6\,keV and $C_{4}=$8--10\,keV bands (see Figure~\ref{4U 1909 Figure 3}(b) and~\ref{4U 1909 Figure 3}(c)).  We note a hint of a spectral softening during the off state, particularly in the $C_{1}=$2--4\,keV and $C_{4}=$8--10\,keV hardness ratio.  No strong variations in the hardness ratios are observed during the flares, where the count rate increased to $\sim$16\,counts s$^{-1}$.  The hardness ratio between $C_{3}=$6--8\,keV and $C_{4}=$8--10\,keV bands shows no significant variations as a function of time (see Figure~\ref{4U 1909 Figure 3}(d)).}

\subsection{Fractional RMS}
\label{Pulsation Strength}
To monitor changes in the strength of the pulsations in each observation {\oldbf as a function of time}, we calculated the fractional root mean square (rms) amplitude and its uncertainty for each interval defined in Section~\ref{Temporal Analysis} using Equations 10 and B2 in \citet{2003MNRAS.345.1271V}, respectively (see Figure~\ref{4U 1909 Figure 6}(a)).  {\oldbf Since the commonly used peak-to-peak pulsed fraction

\begin{equation}
  \label{Modulation Amplitude}
  \mathcal{P}=\frac{F_{\rm max}-F_{\rm min}}{F_{\rm max}+F_{\rm min}}
\end{equation}

\noindent
suffers from systematic uncertainties in calculating the underlying minimum and maximum of the pulse profile (see e.g. Pottschmidt et al. in prep), we chose the fractional rms amplitude to measure the relative amplitude of the observed pulse profile.}  We find that the fractional rms amplitude during Observation I reaches a minimum of (-0.3\,$\pm$\,0.6)$\%$ 3--6\,ks into the observation, which is consistent with zero.  {\oldbf It is important to note that pulsations were still detected for much of this light curve segment apart from a $\sim$602\,s pulse period nearly $\sim$4\,ks into the observation  (see Figure~\ref{4U 1909 Figure 5}).  We calculated the fractional rms of the light curve segment after excising the $\sim$602\,s interval and find it to be (0.5\,$\pm$\,0.4)\,$\%$.  We then observe the fractional rms amplitude increases to (2.00\,$\pm$\,0.08)$\%$ about 9\,ks into the observation corresponding to the {\oldbf flares presented in Section~\ref{Temporal Analysis}}, and then peaks at a maximum of 3.7\,$\pm$\,0.1$\%$ about $\sim$34.5\,ks into the observation.}

In Observation II, we find that a rapid increase in the strength of the pulsations during the first 10\,ks of the observation where the fractional rms amplitude increases from (5.8\,$\pm$\,0.3)\,$\%$ to (8.4\,$\pm$\,0.4)\,$\%${\oldbf .  This is coincident with flaring activity similar to that observed in Observation I.  Towards the end of the observation, we observe the fractional rms amplitude to decrease to less than 3\,$\%$.}  We find no evidence of a pulse dropout; however, the fractional rms in Segment 7 was found to be (1.4\,$\pm$\,0.2)\,$\%$.

\section{Spectral Analysis}
\subsection{Time-Averaged Spectral Analysis}
\label{Averaged Spectral Analysis}
\begin{figure*}[t]
    \centering
    \begin{tabular}{cc}
    \subfigure
    {
        \includegraphics[trim=0cm 0cm 0cm 0cm, clip=false, scale=0.3, angle=0]{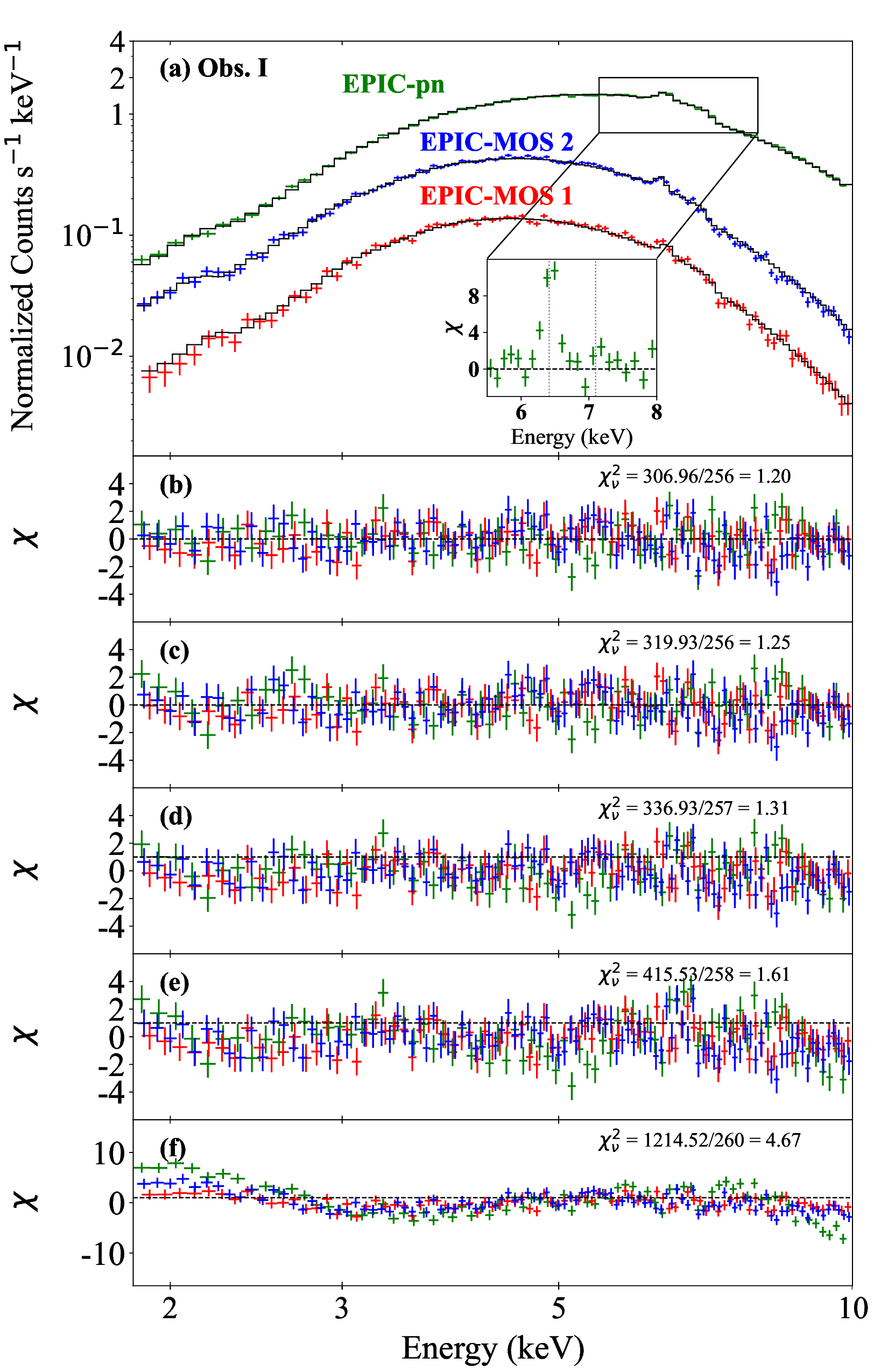}
        \label{4U 1909 Figure 7a}
    }
    &
    \subfigure
    {
        \includegraphics[trim=0cm 0cm 0cm 0cm, clip=false, scale=0.3, angle=0]{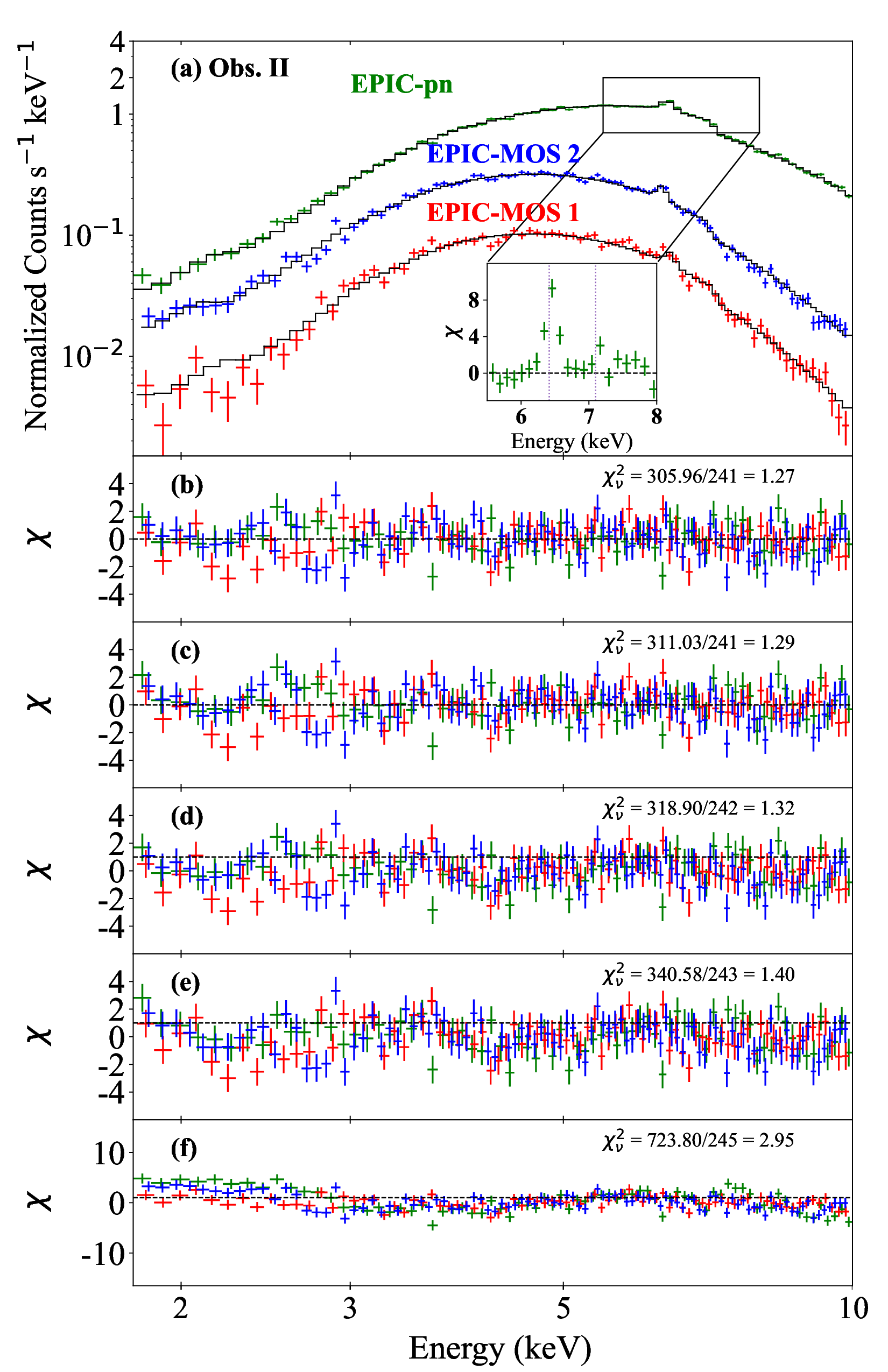}
        \label{4U 1909 Figure 7b}
    }
    \end{tabular}
    \caption{(a) Spectra of 4U 1909$+$07 for Observation I (left) and Observation II (right) where the EPIC-MOS 1, EPIC-MOS 2, and EPIC-pn data are shown in red, blue, and green, respectively.  The best fit \texttt{highecut} model, shown in black, consists of a continuum comprised of an absorbed power law with a high-energy cutoff and emission lines near 6.4\,keV and {\oldbf 7.05\,keV.  The insets show the respective residuals for Fe K$\alpha$ and Fe K$\beta$ without the Gaussian lines.  Spectra are folded with the instrumental response matrices.} (b) Residuals of the best-fit \texttt{highecut} model modified by both fully covered and partially covered absorption columns and the folding energy is allowed to vary.  (c) Residuals of the \texttt{highecut} model modified by a fully covered absorption column and a blackbody.  (d) Residuals where the folding energy is fixed to 26.0\,keV \citep[see e.g.][]{2023ApJ...948...45I}.  (e) Residuals of the \texttt{power} model modified by both fully covered and partially covered absorption column.  (f) Residuals of the \texttt{power} model modified by a single fully covered absorption column.}
    \label{4U 1909 Figure 7}
\end{figure*}

The \xmm\ spectra of 4U 1909$+$07 were analyzed using the package \texttt{XSPEC v12.13.0c}.  We made use of the \texttt{XSPEC} convolution model \texttt{cflux} to calculate the fluxes and associated errors.  To account for instrumental calibration uncertainties, we used cross-calibration constants normalized to EPIC-pn during the spectral analysis (see Table~\ref{Time-Averaged Spectral Parameters}).  We additionally note that for bright sources observed when EPIC-MOS 2 observations are performed in Timing mode, residual calibration uncertainties in the effective area and redistribution are known to produce a systematically softer spectral continuum compared to EPIC-pn \citep{2014A&A...564A..75R}.  This is the case for both Observations I and II, where the spectral slope is found to be steeper in EPIC-MOS 2 compared to EPIC-pn.  To account for this known systematic difference in the spectral shape, we fit an additive gain-shift offset for EPIC-MOS 1 and MOS 2, and fix the offset for EPIC-pn, which we use as a reference, to zero \citep[see][for details on the charge transfer modeling]{2005SPIE.5898..194B}.  Additionally, the slope for each instrument is set to zero.  This ensures that the derived continuum slope is determined by EPIC-pn.

To monitor changes in the 1--10\,keV spectral shape, we extracted \xmm\ EPIC-pn, EPIC-MOS 1 and EPIC-MOS 2 spectra for Observations I and II respectively {\oldbf (see Figure~\ref{4U 1909 Figure 7})}.  We note the presence of strong residuals near $\sim$6.4\,keV and {\oldbf $\sim$7.05\,keV}, which we interpret as the Fe K$\alpha$ and Fe K$\beta$ emission features, as well as an Fe K edge at {\oldbf $\sim$7.12\,keV}.  We tested a power law (\texttt{power}), a power law with a high-energy cutoff \citep[\texttt{highecut};][]{1983ApJ...270..711W}, a cutoff power-law (\texttt{cutoffpl}), and a negative-positive exponential cutoff \citep[\texttt{npex};][]{1999ApJ...525..978M}, which were included in the broadband analysis presented in \citet{2023ApJ...948...45I}.  All models were modified by a single fully covered neutral absorber (\texttt{tbabs} in XSPEC) using the \citet{1996ApJ...465..487V} cross sections and \citet{2000ApJ...542..914W} abundances.  We additionally allowed the abundance of Fe in the absorber to vary to account for the Fe K-shell ionization edge (\texttt{tbfeo} in XSPEC).  We found that the spectral shape is best described with the \texttt{highecut} model with a hard spectral index of {\oldbf 1.21\,$\pm$\,0.08} and {\oldbf 1.02$^{+0.09}_{-0.08}$} for Observations I and II {\oldbf (see Figure~\ref{4U 1909 Figure 7}(b))}, although we found the folding energy could not be constrained due to the limited bandpass of \xmm\ and it may {\oldbf also} systematically affect the accuracy of our photon index values.  {\oldbf Additionally, we} tested the \texttt{cutoffpl}, but note the \texttt{cutoffpl} model yielded wavy residuals resulting in reduced $\chi^2$ values of $\sim$1.37 (257) and $\sim$1.41 (242) for Observations I and II, respectively.  {\oldbf In each observation, for the \texttt{npex} model, the photon index could not be constrained.}  The \texttt{highecut} model was preferred over the power law due to significant residuals at high energies, suggesting a spectral turnover (see Figure~\ref{4U 1909 Figure 7}(d)).
    
{\oldbf The additive gain shift offsets for EPIC-MOS 1 and EPIC-MOS 2 are found to be 41$^{+15}_{-16}$\,eV and -54$^{+8}_{-11}$\,eV for Observation I.  For Observation II, the gain shift offset for MOS 1 and MOS 2 are found to be 43$^{+25}_{-26}$\,eV and -79$^{+17}_{-12}$\,eV, respectively (see Table~\ref{Time-Averaged Spectral Parameters}).  We also attempted to fit the \xmm\ spectra where the EPIC-pn offset was free and found that while the $\chi_\nu^{2}$ improves to 1.15 for 255 d.o.f. for Observation I and 1.21 for 240 d.o.f. for Observation II, it did not significantly affect the best-fit continuum parameters and their uncertainties.  We chose to fix the EPIC-pn gain shift offset to zero so that calibration offsets in the MOS 1 and MOS 2 instruments are relative to EPIC-pn.}
  
{\oldbf
\begingroup

\setlength{\tabcolsep}{8pt} 
\renewcommand{\arraystretch}{1.5} 

\begin{table*}
  \caption{Time-Averaged X-ray spectral parameters}
  \label{Time-Averaged Spectral Parameters}

  \centering
  \begin{tabular}{ccc}
    \hline\hline
    Model Parameter & Observation I & Observation II \\
    \hline
    $\chi_\nu^{2}$ (dof) & 1.20 (256) & 1.27 (241) \\
    $C_{\rm EPIC,MOS 1}^{a}$ & 1.08\,$\pm$\,0.01 & 1.10\,$\pm$\,0.02 \\
    $C_{\rm EPIC,MOS 2}^{a}$ & 1.089\,$\pm$\,0.008 & 1.06\,$\pm$\,0.01 \\
    Cutoff Energy (keV) & 7.6$^{+0.3}_{-0.5}$ & 6.1$^{+0.4}_{-0.3}$ \\
    Folding Energy (keV) & $<$12.4$^{b}$ & $<$14.4$^{b}$ \\
    $N_{\rm H,1}$ ($\times$10$^{22}$\,cm$^{-2}$) & 8$^{+2}_{-1}$ & 8$^{+3}_{-1}$ \\
    $N_{\rm H,2}$ ($\times$10$^{22}$\,cm$^{-2}$) & 18.3$^{+0.9}_{-0.8}$ & 21\,$\pm$\,1 \\
    Cvr & 0.94$^{+0.02}_{-0.03}$ & 0.95$^{+0.02}_{-0.03}$ \\
    $\Gamma$ & 1.21\,$\pm$\,0.08 & 1.02$^{+0.09}_{-0.08}$ \\
    \hline
    Fe K$\alpha$ Energy (keV) & 6.41\,$\pm$\,0.01 & 6.41\,$\pm$\,0.02 \\
    Fe K$\alpha$ Width ($\sigma_{\rm Fe K\alpha}$)$^c$ & 0.01 & 0.01 \\
    Fe K$\alpha$ EQW (eV) & 40$^{+7}_{-8}$ & 48\,$\pm$\,8 \\
    Fe K$\alpha$ Flux ($\times$10$^{-5}$\,ph cm$^{-2}$ s$^{-1}$) & 0.9\,$\pm$\,0.1 & 1.0\,$\pm$\,0.1 \\
    \hline
    Fe K$\beta$ Energy (keV)$^e$ & 7.09 & 7.09 \\
    Fe K$\beta$ Width ($\sigma_{\rm Fe K\beta}$)$^c$ & 0.01 & 0.01 \\
    Fe K$\beta$ EQW (eV) & 12\,$\pm$\,6 & 21$^{+9}_{-8}$ \\
    Fe K$\beta$ Flux ($\times$10$^{-5}$\,ph cm$^{-2}$ s$^{-1}$) & 0.22\,$\pm$\,0.07 & 0.21\,$\pm$\,0.07 \\
    \hline
    CS Flux ($\times$10$^{-5}$\,ph cm$^{-2}$ s$^{-1}$) & $<$4.0 & $<$5.1 \\
    \hline
    Fe K$\beta$/K$\alpha$ & 0.24\,$\pm$\,0.08 & 0.22\,$\pm$\,0.08 \\
    Fe abundance$^{f}$ & 1.0\,$\pm$\,0.2  & 0.9\,$\pm$\,0.1 \\
    \hline
    $GS_{\rm EPIC,MOS 1}^{g}$ & 41$^{+15}_{-16}$ & 43$^{+25}_{-26}$ \\
    $GS_{\rm EPIC,MOS 2}^{g}$ & -54$^{+8}_{-11}$ & -79$^{+17}_{-12}$ \\
    \hline
    Flux ($\times$10$^{-10}$\,erg cm$^{-2}$ s$^{-1}$)$^h$ & 2.3\,$\pm$\,0.1 & 1.77$^{+0.09}_{-0.06}$ \\
    \hline
  \end{tabular}
  \tablecomments{\\*
    $^a$ Detector cross-calibration constants with respect to EPIC-pn. \\*
    $^b$ The upper limit of the folding energy $E_{\rm fold}$ in keV. \\*
    $^c$ The width of the Fe K$\alpha$ and Fe K$\beta$ emission lines are frozen to 0.01\,keV. \\*
    $^d$ Flux for a step function where the step line energies are tied to the Fe K$\alpha$ flux and Fe K$\alpha-$0.16 with a width of both fixed at zero. \\*
    $^e$ The energy is frozen because we can only obtain an upper limit. \\* 
    $^f$ Relative to ISM abundances. \\*
    $^g$ Additive gain shift in eV {\oldbf with respect to EPIC-pn}. \\*
    $^h$ Unabsorbed flux in the 1--10\,keV band.}

\end{table*}
\endgroup
}

To improve our constraints on the photon index, we initially attempted to fix the folding energy to 26\,keV, which was found in broadband analyses using archival \textsl{NuSTAR} observations \citep{2020MNRAS.498.4830J,2023ApJ...948...45I}.  We found{\oldbf ,} however, that this results in an overestimation of the \xmm\ data at high energies and a reduced $\chi^{2}$ of {\oldbf 1.31} for {\oldbf 257} degrees of freedom and {\oldbf 1.32} for {\oldbf 242} degrees of freedom for Observations I and II, respectively (see Figure~\ref{4U 1909 Figure 7}(c)).  To best fit the data, we chose to allow the folding energy to vary within the range of 10--40\,keV.  We found we could not constrain the folding energy and instead estimated its upper limit to be {\oldbf 12.4\,keV} and {\oldbf 14.4\,keV} in Observations I and II, respectively.  This is not surprising since this is outside of the energy bandpass of \xmm.

In addition to the high-energy cutoff, our initial fits of the spectra for both observations showed significant positive residuals below $\sim$3\,keV (see Figure~\ref{4U 1909 Figure 7}(e)).  As in earlier studies of 4U 1909$+$07 using combined \rxte\ and \integral\ observations \citep{2011A&A...525A..73F} or with \suzaku\ observations \citep{2012A&A...547A...2F,2013ApJ...779...54J}, we accounted for the soft excess either using an additional partial covering absorption convolution model

%
\begin{equation}
	\mathtt{partcov}(E) = (1-f_{\text{pcf}})*\mathtt{tbfeo}_{1}(E) + f_{\text{pcf}}*\mathtt{tbfeo}_{2}(E),
  \label{eqn:partcov}
\end{equation}
\noindent
or a thermal blackbody (\texttt{bbodyrad}).  We note that \citet{2012A&A...547A...2F} reported different best-fit temperatures of the thermal blackbody using \textsl{Suzaku} depending on the model that was used.  For the partial covering absorption column, we found the neutral hydrogen column density for Observations I and II to be {\oldbf (18.3$^{+0.9}_{-0.8}$)$\times$10$^{22}$\,cm$^{-2}$} and {\oldbf (21\,$\pm$\,1)$\times$10$^{22}$\,cm$^{-2}$} and the covering fraction to be 0.94$^{+0.02}_{-0.03}$ and 0.95$^{+0.02}_{-0.03}$, respectively (see Table~\ref{Time-Averaged Spectral Parameters}).  For completeness, we alternatively used a thermal blackbody component to account for the soft excess, which reduces our $\chi^{2}_\nu$ to {\oldbf 1.25 for 256} dof and {\oldbf 1.29 for 241} dof for Observations I and II, respectively {\oldbf (see Figure~\ref{4U 1909 Figure 7}(c))}.  In Observation I, we found the temperature and radius of the blackbody to be {\oldbf 0.19\,$\pm$\,0.02\,keV} and {\oldbf 250\,$\pm$\,77\,km}, respectively.  The temperature and radius of the blackbody in Observation II were found to be {\oldbf 0.20\,$\pm$\,0.02\,keV} and {\oldbf 207\,$\pm$\,66\,km}, respectively.  We prefer the partial covering absorber to describe the soft excess as it resulted in slightly improved fit statistics compared to the blackbody emission component (see Figure~\ref{4U 1909 Figure 7}) and that the radius of the thermal blackbody is more consistent with that of an accretion disk around the neutron star, as opposed to a hot spot on the compact object, as what would be expected for wind-fed SGXBs.

In the inset in Figure~\ref{4U 1909 Figure 7}, we plot the spectrum from 5.5--8.0\,keV, which shows clear Fe K$\alpha$ and Fe K$\beta$ emission lines.  The flux ratios of the Fe K$\beta$ line relative to the Fe K$\alpha$ line were found to be {\oldbf 0.24\,$\pm$0.08} and {\oldbf 0.22\,$\pm$\,0.08} for Observations I and II, which is {\oldbf slightly larger than} the value of 0.13 expected for fluorescence \citep{2003A&A...410..359P}.

We note that \citet{2010ApJ...715..947T} found a strong Compton shoulder at an energy of $\sim$6.34\,keV with a flux ratio {\oldbf of} 40$\%$ to the main Fe K$\alpha$ line using \chandra\ HETG observations.  To investigate the possible presence of a Compton shoulder in our \xmm\ observations, we initially added a Gaussian emission line at an energy frozen to 6.34\,keV with a width frozen to 0.01\,keV.  We find the {\oldbf additional Gaussian component} does not significantly improve the fit quality, where the $\chi^2/$d.o.f. changes from {\oldbf 295.36/257 to 293.28/256 during Observation I and 296.79/242 to 294.97/241 during Observation II}.  We determined the false alarm probability of the Compton shoulder using the Monte Carlo analysis described in \citet{2002ApJ...571..545P} to be {\oldbf $\sim$14.4$\%$ and $\sim$16.2$\%$} in Observations I and II with 10$^{4}$ trials, respectively.  We find an upper limit of {\oldbf $\sim$4.0$\times$10$^{-5}$\,photons cm$^{-2}$ s$^{-1}$ and $\sim$5.9$\times$10$^{-5}$\,photons cm$^{-2}$ s$^{-1}$} for the flux of the Compton shoulder during Observations I and II, respectively.

Another commonly used model to account for an asymmetry on the red wing of the Fe K$\alpha$ line is two-step function convolved with a Gaussian, \citep[e.g. OAO 1657-415;][]{2019MNRAS.483.5687P}.  To search for a possible Compton shoulder, we tied one of the line energies of the two-step function to the energy of the primary Fe K$\alpha$ line and the second is fixed at an energy 0.16\,keV less than the primary line

\begin{equation}
  \label{Compton Shoulder}
  E_{1}=K_\alpha-0.16\,{\rm keV,}
\end{equation}
\noindent
with the widths fixed to zero{\oldbf ,} and the normalizations are equal, but opposite in sign {\oldbf \citep[see e.g.][]{2002MNRAS.337..147M,2003ApJ...597L..37W}}.  {\oldbf Again, a Monte Carlo analysis was performed,} but for the two step function we removed both components of the step function as the null hypothesis.  We find the false alarm probability of the Compton shoulder using the Monte Carlo analysis described in \citet{2002ApJ...571..545P} to be 8.1$\%$ and 14.2$\%$ for Observations I and II with 10$^{4}$ trials, respectively.  We find an upper limit of {\oldbf $\sim$4.0$\times$10$^{-5}$\,photons cm$^{-2}$ s$^{-1}$ and $\sim$5.0$\times$10$^{-5}$\,photons cm$^{-2}$ s$^{-1}$} for the flux of the Compton shoulder during Observations I and II, respectively.

\begin{figure*}[t]
    \centering
    \begin{tabular}{cc}
    \subfigure
    {
        \includegraphics[trim=0cm 0cm 0cm 0cm, clip=false, scale=0.3, angle=0]{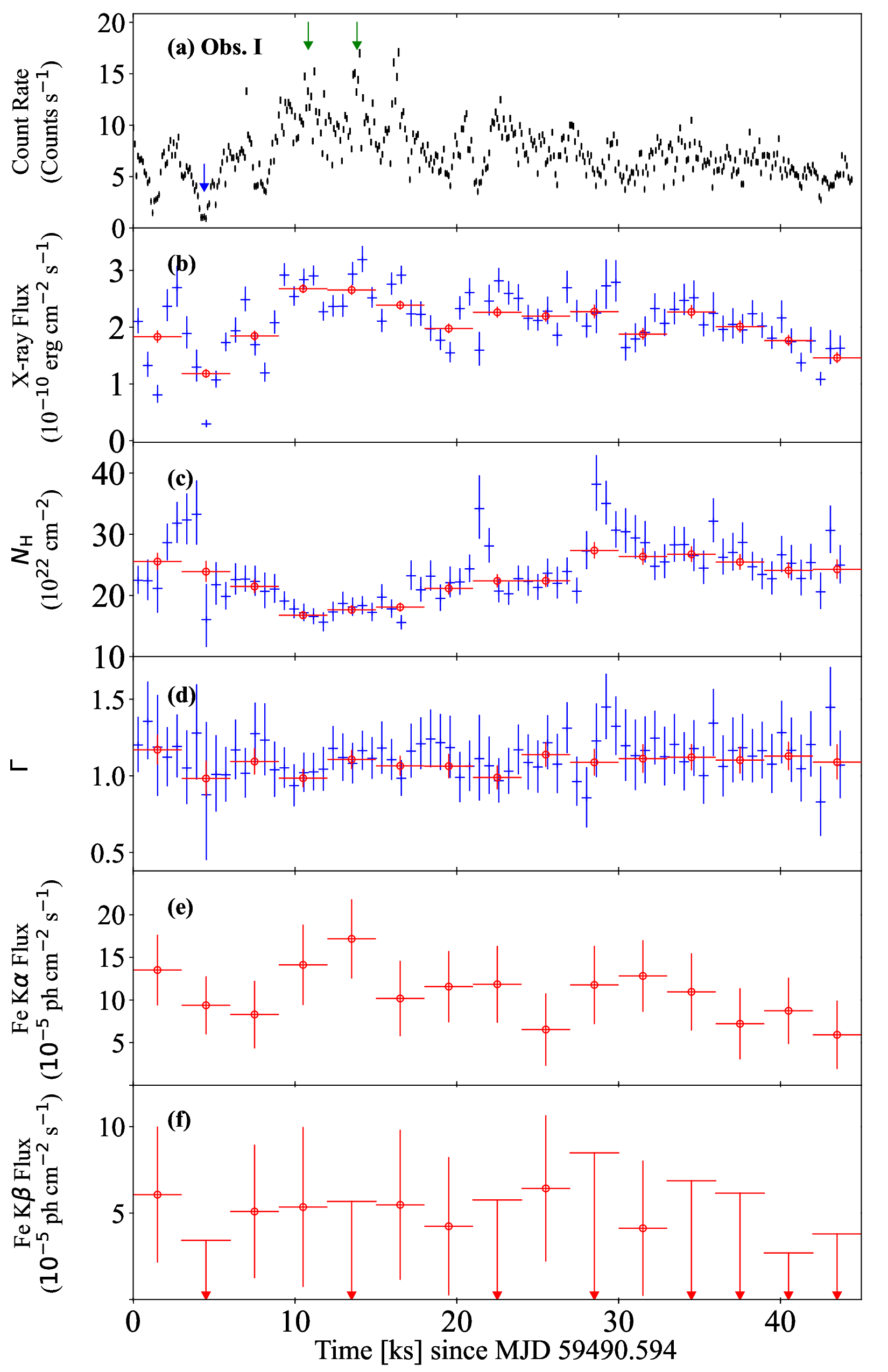}
        \label{4U 1909 Figure 8a}
    }
    &
    \subfigure
    {
        \includegraphics[trim=0cm 0cm 0cm 0cm, clip=false, scale=0.3, angle=0]{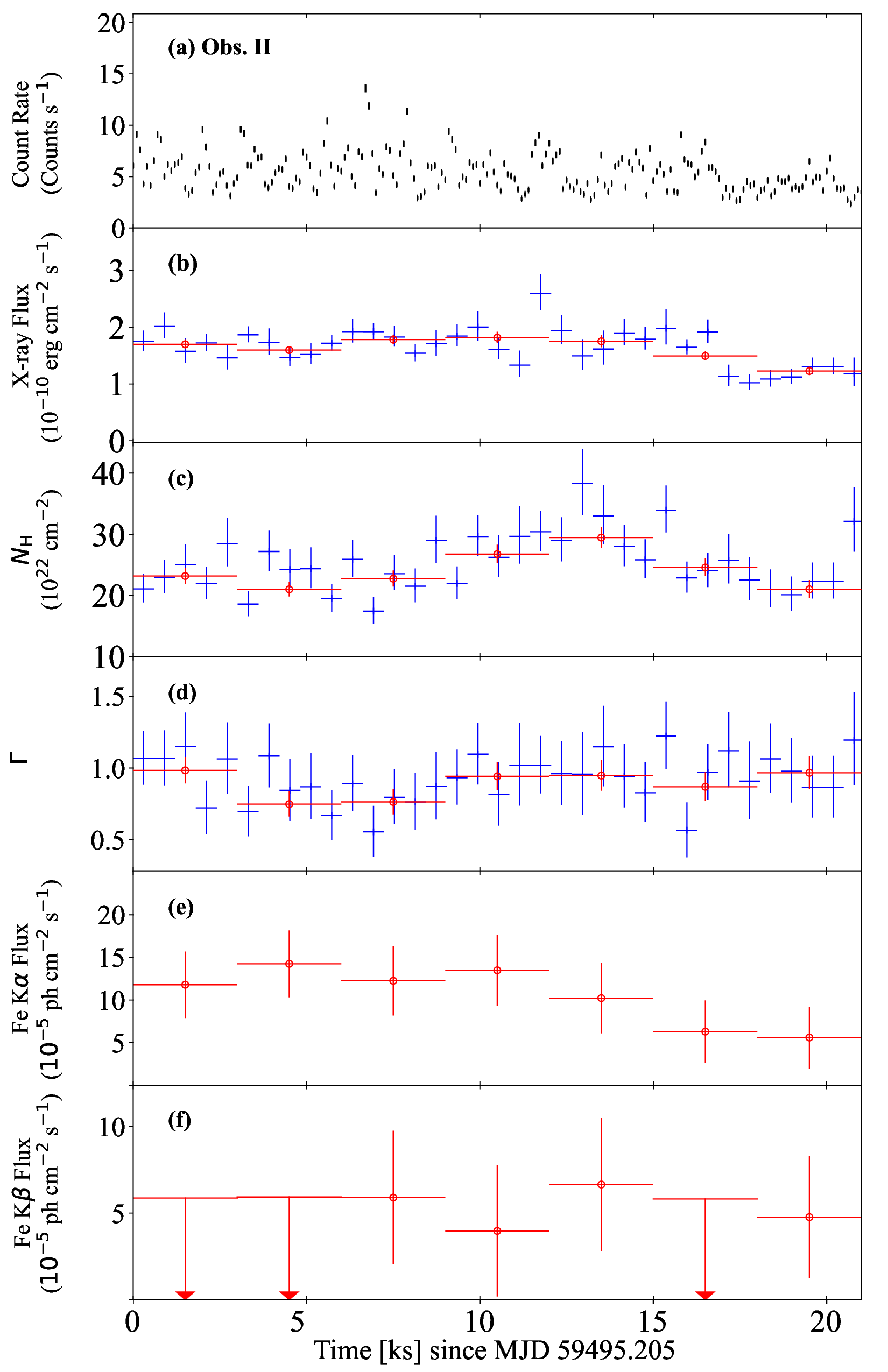}
        \label{4U 1909 Figure 8b}
    }
    \end{tabular}
    \caption{Spectral parameters as a function of time using the \texttt{highecut} model.  The 1--10\,keV EPIC-pn light curve for observations I (left) and II (right) is shown in black (panel a).  The times of the off-state and flares are indicated by the blue and green arrows, respectively.  The ks-integrated and pulse-to-pulse analyses are indicated in red and blue, respectively.}
    \label{4U 1909 Figure 8}
\end{figure*}

{\oldbf \subsection{ks-Integrated Spectral Analysis}
\label{sec:ks-integrated}}

Since the hardness ratios were observed to show significant variations with time {\oldbf (see Figure~\ref{4U 1909 Figure 3}(b)-(c)-(d))}, we investigated the evolution of the spectral shape as a function of time using time-resolved spectroscopy.  We initially chose to subdivide the light curves for each observation using the 3\,ks time intervals defined in Section~\ref{Temporal Analysis}, which we refer to as our ks-integrated spectral analysis.

In Figure~\ref{4U 1909 Figure 8}, we show the EPIC-pn light curves as well as the unabsorbed 1--10\,keV X-ray flux, local absorption column density $N_{\rm H}$, photon index, and the fluxes of the Fe K$\alpha$ and Fe K$\beta$ emission lines derived from the \texttt{highecut} model at different time bins for Observations I and II.  We note that the best-fit model does not require an additional partial covering absorber, although it is required in the time-averaged spectra (see Table~\ref{Time-Averaged Spectral Parameters}).  We also tested the addition of a partial covering absorber, but found that the covering fraction of the second absorber approaches unity.  We, therefore, confirmed a single absorber yields an accurate description of the data.  While we initially allowed the continuum spectral components described in Table~\ref{Time-Averaged Spectral Parameters} to be free parameters and performed spectral fits on each of 3\,ks intervals, we found due to the reduced signal to noise ratio and the limited 1--10\,keV bandpass of \xmm\ that the cutoff and folding energies were not always constrained.  We chose to fix the cutoff and folding energies to their time-averaged values.  {\oldbf We find the photon index remains constant in both observations (see Figure~\ref{4U 1909 Figure 8}(d)), which suggests the observed variability is likely driven by changes in the absorber (see Figures~\ref{4U 1909 Figure 3}(b) and~\ref{4U 1909 Figure 8}(d)).  This is further supported by the hardness ratio between 6--8\,keV and 8--10\,keV, which shows little variation with time.}

We find the absorption column shows remarkably different behavior between Observations I and II (see Figure~\ref{4U 1909 Figure 8}(c)).  In Observation I, {\oldbf we initially observe the column density to be $\sim$2.4$\times$10$^{23}$\,cm$^{-2}$, which then begins to decrease during the rise of the flare.  Near the peak of the flare, we observe that $N_{\rm H}$ reaches a minimum of $\sim$1.7$\times$10$^{23}$\,cm$^{-2}$.}  It subsequently increases during the decay of the flare to {\oldbf an average value of $\sim$2.5$\times$10$^{23}$\,cm$^{-2}$, which is similar in value to that observed before the onset of the flare.}  In Observation II, the neutral hydrogen column density shows strong variability reaching a maximum of $\sim$2.7$\times$10$^{23}$\,cm$^{-2}$ without a corresponding change in flux.

While the variability of the Fe K$\alpha$ emission feature is not as strong as that observed for the neutral hydrogen column density, moderate variability in the line flux is observed.  We find the Fe K$\alpha$ flux appears to track the continuum in Observation II, but the variability in our first observation appears to be marginal (see Figure~\ref{4U 1909 Figure 8}(e)).  For the Fe K$\beta$ emission feature, we find that it is not strongly detected at most time intervals.  We instead measure the upper limit to be (5--10)$\times$10$^{-5}$\,photons cm$^{-2}$ s$^{-1}$  (see Figure~\ref{4U 1909 Figure 8}(f)).

\subsection{Pulse-to-Pulse Spectral Analysis}
\label{Pulse-to-Pulse Spectral Analysis}
To further investigate variations in the absorption column and spectral shape as a function of time, we additionally extracted a 1--10\,keV spectrum for each $\sim$602.6\,s rotation period of the neutron star, resulting in 73 individual spectra for Observation I and 36 individual spectra for Observation II.  We additionally note that an analysis on even shorter timescales is not sensible due to strong changes in the spectral shape over the neutron star rotation period \citep[see e.g.][]{2012A&A...547A...2F,2023ApJ...948...45I}.  Due to the reduced S/N, we only considered the EPIC-pn spectra since the low exposure of the spectra precluded accurate constraints of the spectra using the EPIC-MOS 1/2 instruments.  While we initially attempted to measure the flux of the Fe K$\alpha$ emission feature on a pulse-to-pulse basis, we found it cannot be constrained.  So, we fixed the fluxes of the Fe K$\alpha$ line to the values in our coarser ks-integrated analysis presented in Section~\ref{sec:ks-integrated}.  This resulted only in an analysis of the variability of the 1--10\,keV flux, the neutral hydrogen absorption column density and the photon index.  We note that no soft excess was observed and we, therefore, chose to model the absorption column with a single fully-covered absorber where the Fe abundance was frozen to the time-averaged values.

In {\oldbf Figure~\ref{4U 1909 Figure 8}(b)-(c)-(d)}, we plot the pulse-to-pulse results for the 1--10\,keV flux, absorption column density and photon index using EPIC-pn.  The average $\chi^{2}_\nu$ for Observations I and II are 1.10 and 1.14, respectively.  The standard deviation of $\chi^{2}_\nu$ are 0.23 and 0.36 for the first and second observations, respectively.  We find the photon index in both observations to be stable including the observed pulse dropout {\oldbf (see Figure~\ref{4U 1909 Figure 8}(d)), further supporting} that the observed variability is due to changes in the {\oldbf neutral} hydrogen absorption column.  Similar to our 3\,ks resolution spectroscopy, we find strong variability in the neutral hydrogen absorption column density in both observations{\oldbf ,} but with much more local instability including a sudden drop in $N_{\rm H}$ during the pulse dropout in the first observation  (see Figure~\ref{4U 1909 Figure 8}(c)).  Increases in the absorption column are coincident with the other dips in the each observation, which suggests that these are driven by clumps in the wind of the B-type supergiant star along the line of sight. The absorption column tracks the hardness ratio in the 2--4\,keV and 8--10\,keV bands, which suggests that the observed spectral hardness is tied to the changes in the intervening material along the line of sight.  We additionally found episodes of low absorption, which are coincident with flares.  This could possibly be explained by the accretion of clumps along the line of sight.

\section{Discussion}
\label{Discussion}

\subsection{Geometry of the Absorbing Medium}
\label{Geometry of the Absorbing Medium}
From the pulse-to-pulse analysis of the first \xmm\ observation of 4U 1909$+$07, we observed the minimum and maximum X-ray fluxes to be {\oldbf (2.9\,$\pm$\,0.7)$\times$10$^{-11}$\,erg cm$^{-2}$ s$^{-1}$} and (3.2\,$\pm$\,0.2)$\times$10$^{-10}$\,erg cm$^{-2}$ s$^{-1}$, respectively (see Figure~\ref{4U 1909 Figure 8}(b)).  Calculating the peak-to-trough flux ratio, we find the dynamic range to be an order of magnitude, which is consistent with that typically found in classical SGXBs.  This is further supported by the minimum and maximum X-ray fluxes in the second observation.  These were observed to be (1.0\,$\pm$\,0.1)$\times$10$^{-10}$\,erg cm$^{-2}$ s$^{-1}$ and {\oldbf (2.6\,$\pm$\,0.3)$\times$10$^{-10}$\,erg cm$^{-2}$ s$^{-1}$}, respectively, resulting in a peak-to-trough flux ratio of {\oldbf 2.6}.  Larger values in the peak-to-trough intensity ratio up to 10$^{6}$ are observed in prototypical Supergiant Fast X-ray Transients (SFXTs) where the donor stars are also OB supergiants \citep[e.g. IGR J17544-2619; IGR J16479-4514,][]{2015A&A...576L...4R,2020ApJ...900...22S}.  Possible explanations to account for these flux differences between SGXBs and SFXTs include the quasi-spherical settling accretion regime \citep[see e.g. IGR J11215-5952,][]{2017ApJ...838..133S}, co-rotating interaction regions \citep[CIRs,][]{2017A&A...608A.128B}{\oldbf ,} or a centrifugal barrier \citep{2007AstL...33..149G}.

The X-ray spectral energy distribution of 4U 1909$+$07 can be described by a hard power law with a spectral index of $\Gamma\sim$1, a high-energy cutoff and a partial covering absorption component (see Table~\ref{Time-Averaged Spectral Parameters}), similar to that observed in other classical SGXBs \citep[see e.g. Vela X-1,][]{2014ApJ...780..133F}.  It is important to note, however, that while a partial covering absorber was required in the time-averaged spectral fits, we found no evidence of it in the time-resolved spectral analysis--neither in our ks-integrated results (see Section~\ref{sec:ks-integrated}){\oldbf ,} nor our pulse-to-pulse analysis (see Section~\ref{Pulse-to-Pulse Spectral Analysis}).  This is evidence that the observed local absorption column density in the time-averaged spectra may be driven by a superposition of different absorption column densities from the stellar wind of the OB supergiant star \citep[see e.g. 4U 1538-522,][]{2014ApJ...792...14H}.  {\oldbf As a result, the average absorption column density can be described by a partial coverer with a high covering fraction due to the superposition of different absorption columns along the line of sight \citep{2016A&A...588A.100M}.}


In wind-fed classical SGXBs, accretion onto the compact object is mediated by the inhomogeneous, or ``clumpy'', stellar wind of the OB supergiant star \citep[see e.g.][for a review]{2017SSRv..212...59M}, although in some cases a transient accretion disk may also be present \citep[see e.g.][]{2019A&A...622A.189E}.  The resulting changes in the mass accretion rate in addition to absorption and/or scattering of X-rays lead to rapid variability in the X-ray flux and spectral shape on timescales of hundreds of seconds to a few kiloseconds.  This fast variability in 4U 1909$+$07 due to the wind of the B0-3 I star is supported by the variations seen in the energy-resolved light curves and corresponding hardness ratio between the 2--4\,keV and 8--10\,keV energy bands (see Figure~\ref{4U 1909 Figure 3}(b)), which show strong changes in spectral hardness with time.  Additional evidence of this is observed in our ks-integrated and pulse-to-pulse resolved analyses, where the local absorption column density correlates with the hardness ratio of the 2--4\,keV and 8--10\,keV bands \citep{2023ApJ...945...51P}.  It is important to note{\oldbf ,} however, that the hardness ratio of the high-energy bands is constant, which may be evidence that the continuum emission from the neutron star is stable across the observation \citep[see e.g. Vela X-1,][]{2023A&A...674A.147D}.

Changes in the X-ray luminosity due to accretion can also ionize the OB stellar wind, which results in changes in its density and/or velocity.  In each of our \xmm\ observations, we find a clear decrease in the local absorption column density during the peaks of the flares (see Figure~\ref{4U 1909 Figure 8}(c)), followed by a rapid increase in the column density as the flare decays.  This is possible evidence that the higher flux during the peak of the flare can photoionize the surrounding material from which the neutron star is accreting, which has been observed in both SGXBs \citep[e.g. OAO 1657-415;][]{2023ApJ...945...51P} and SFXTs \citep{2017A&A...608A.128B}.  Similarly, the increase of $N_{\rm H}$ following the flare may be driven by recombination in the wind material as the X-ray flux decreases.

The wind structure observed in SGXBs is additionally highly structured due to the line-driven instability and the influence of the gravitational field of the pulsar.  This commonly results in phase-locked structures such as accretion wakes, which cross our line of sight and are observable at late orbital phases \citep[e.g. Vela X-1; 4U 1538-522,][]{2015A&A...577A.130F,2023A&A...674A.147D}.  Additionally, photoionization wakes may be present where the complex wind interacts with the Str\"omgren sphere created by the photoionization of the wind material.  If an accretion or photoionization wake is present in the wind of the OB supergiant as it interacts with the neutron star, a phase-locked increase in the absorption column density would be observed that is consistently present at late orbital phases.  While our \xmm\ campaign of 4U 1909$+$07 spans orbital phases 0.664–0.781 where the average local absorption column density peaks on the order of $\sim$2$\times$10$^{23}$\,cm$^{-2}$, we do not observe a phase-locked rapid increase in $N_{\rm H}$ that would be suggestive of an accretion wake along our line of sight.  This may result from the orbital inclination of 4U 1909$+$07, which has been observed to be 46--58$^{\circ}$ \citep{2015A&A...578A.107M}.


In addition to rapid changes in the X-ray flux and spectral continuum, both of our \xmm\ observations showed the presence of strong Fe K$\alpha$ and K$\beta$ emission lines in addition to a strong Fe Kedge, which are typically present in SGXBs likely due to the extended stellar wind or reflection off the stellar photosphere.  These emission and absorption features have been previously found in 4U 1909$+$07 with \chandra\ HETG \citep{2010ApJ...715..947T}, \suzaku\ XIS \citep{2012A&A...547A...2F,2013ApJ...779...54J} and \nustar\ \citep{2020MNRAS.498.4830J,2023ApJ...948...45I}. It is important to note that in our \xmm\ campaign, the flux ratios of the Fe K$\beta$ line relative to the Fe K$\alpha$ line were found to be consistent with the value of 0.13 expected for fluorescence \citep{2003A&A...410..359P}.  This is in contrast with the higher flux ratios of $\sim$0.3 that were previously observed with \chandra, \suzaku, and \nustar.  One possible explanation is that earlier measurements of the Fe K$\beta$ line may be hampered due to the overlapping Fe Kedge, which was modeled in earlier works using an absorption feature as opposed to a free Fe abundance.  If we instead fix the Fe abundance to 1 and model the Fe Kedge with an absorption edge feature (\texttt{edge} in XSPEC), we also find the flux ratio of the Fe K$\beta$ line relative to the Fe K$\alpha$ line to be 0.3.  However, we prefer to model the Fe Kedge using a variable Fe abundance because of the slight improvement in the {\oldbf test} statistics.

In their survey of 10 High-mass X-ray binaries and 31 Low-mass X-ray binaries using the \chandra\ HETG, \citet{2010ApJ...715..947T} found that the equivalent width Fe K$\alpha$ linearly correlates with the intrinsic column density of the absorbing medium.  They measured the slope to be (3.29\,$\pm$\,0.05)$\times$10$^{22}$\,eV$^{-1}$\,cm$^{-2}$, which is consistent with theoretical predictions presented in \citet{2004ApJS..155..675K} under the assumption of a spherical geometry and solar abundances.  It is important to note that for 4U 1909$+$07{\oldbf ,} \citet{2010ApJ...715..947T} observed an anticorrelation between the Fe K$\alpha$ equivalent width and the intrinsic column density{\oldbf .  This} is in variance with our observations with \xmm\ where the Fe K$\alpha$ equivalent width and the intrinsic column density are observed to be correlated.  One possible explanation may be driven by our viewing geometry of the Fe fluorescent region where the \chandra\ observations were performed at orbital phases $\sim$0.48, $\sim$0.59, and $\sim$0.97 and the \xmm\ observations were performed at phases 0.664--0.781 and 0.712--0.769.

\subsection{Pulse Profile Evolution}
The shape of the pulse profiles in accreting X-ray pulsars has been found to depend on the emission processes and the relative contribution of the two accretion columns \citep[][Falkner et al. in prep.]{1985ApJ...299..138M,1989ESASP.296..433K}.  Our timing analysis of 4U 1909$+$07 has shown that while the overall shape of the pulse profile is similar between Observations I and II{\oldbf , the relative strength of the components between pulse phases 0.0--0.3 and 0.6--0.9 shows significant differences.  We observe that the}  broad structure evolves from a relatively flat plateau in Observation I to a secondary peak in Observation II (see Figure~\ref{4U 1909 Figure 4}).  Our pulse profile maps for each observation shows similar changes in the shape of the broad structure on timescales as short as a few ks (see Figure~\ref{4U 1909 Figure 6}(c)).  The narrow emission peak was also found in earlier \integral\ \citep{2011A&A...525A..73F}, \suzaku\ \citep{2012A&A...547A...2F,2013ApJ...779...54J}, \nustar\ \citep{2020MNRAS.498.4830J,2023ApJ...948...45I}, and \astrosat\ \citep{2020MNRAS.498.4830J} observations, but marginal changes in the secondary structure were found.

The predominantly consistent shape of the pulse profiles in our \xmm\ observations in comparison with earlier results suggest a possible persistent emission geometry of the pulsar at different orbital and superorbital phases.  The changes in the morphology of the pulse profile during the broad structure in between the narrow peaks may indicate changes in the local variation in the column density and distribution of the inhomogeneous wind of the early-type supergiant star \citep{2020MNRAS.498.4830J}.  In some accreting X-ray pulsars, changes in the local variation of the absorbing material result in absorption dips at defined pulse phases of the pulsar.  This has been observed in Be X-ray binaries during outbursts \citep[e.g. 1A 1118-61, EXO 2030$+$375;][]{2012MNRAS.420.2307M,2017A&A...606A..89F,2024A&A...688A.213T,2024A&A...688A.214B}.

{\oldbf Additionally, our} \xmm\ campaign shows a dramatic evolution in the strength of the pulse profile in both observations, including an off state where the $\sim$602\,s pulse period is not detected.  To investigate changes in the pulse strength as a function of the X-ray flux, we plot the fractional rms of the pulse with the unabsorbed 1--10\,keV X-ray flux (see Figure 9).  In Observation II, we observe a bimodal behavior between the fractional rms of the pulse profile and the unabsorbed 1--10\,keV flux, {\oldbf which is} similar to the behavior of the fractional rms compared to the intrinsic neutral hydrogen column density.  While we observe some scatter in Observation I, we found evidence of a correlation between the fractional rms of the pulse profile and the unabsorbed 1--10\,keV flux.

The strength of the pulse profiles in some accreting X-ray pulsars has also been tied to obscuration effects due to an optically thick medium such as an accretion disk \citep[e.g. Her X-1; SMC X-1,][]{2002astro.ph..3213K,2019ApJ...875..144P} or clumpy wind, while in other sources the pulse amplitude has been linked to changes in the mass accretion rate \citep[e.g. EXO 2030$+$375,][]{2017A&A...606A..89F,2021JApA...42...33J}.  Our \xmm\ observations suggest that the fractional rms amplitude of the pulse profile of 4U 1909$+$07 depends on the unabsorbed 1--10\,keV X-ray flux, where it reaches a minimum of (-0.3\,$\pm$\,0.6)$\%$ at an X-ray flux of (1.11\,$\pm$\,0.07)$\times$10$^{-10}$\,erg cm$^{-2}$ s$^{-1}$.  Since X-ray flux is directly proportional to the mass accretion rate, we propose that the observed pulse dropout is driven by a reduced mass accretion rate (see Figure 9) compared to changes in the intrinsic absorption of the source.

\subsection{Comparison with Vela X-1}
To place our observations of the pulse dropout in 4U 1909$+$07 in context, we compare it with Vela X-1, which is the prototypical wind-fed SGXB where mass transfer is mediated directly by wind accretion.  Vela X-1 consists of a {\oldbf massive 1.8--2.1\,$M_\odot$} neutron star rotating at $\sim$283\,s \citep{1976ApJ...206L..99M}{\oldbf ,} and a B0.5 Ib supergiant donor star HD 77523 \citep{1972ApJ...175L..19H} {\oldbf with a mass of 21--28\,$M_\odot$} orbiting their center of mass with a period of $\sim$8.964\,days.  While the average luminosity in Vela X-1 was found to be 5$\times$10$^{36}$\,erg s$^{-1}$ \citep{2010A&A...519A..37F}, its flux is strongly variable, ranging from off states where the observed flux decreases to less than 10$\%$ of its normal value to bright, giant flares where the 20--40\,keV flux reached a peak of 7\,Crab \citep{2008A&A...492..511K,2022A&A...660A..19D}.  The off states and flaring behavior is similar to{\oldbf ,} but more extreme compared to our \xmm\ campaign of 4U 1909$+$07 (see Section~\ref{Timing Analysis}){\oldbf , possibly due to differences in the mass loss rate and wind speed between the B0.5 Ib star in Vela X-1 and the B0-3 Ib star in 4U 1909$+$07.}

We first discuss the observed pulse dropout in 4U 1909$+$07 and how it compares with similar off states in Vela X-1.  During an uninterrupted campaign using \integral\ {\oldbf ISGRI} that spanned two weeks,  \citet{2008A&A...492..511K} identified a total of five separate{\oldbf ,} but closely spaced off states{\oldbf ,} ranging from $\sim$520\,s to $\sim$1980\,s in duration, {\oldbf and} all but the last off state showed no identifiable transition periods between the start and end of the dip.  From the \integral\ observations, \citet{2008A&A...492..511K} observed the {\oldbf 20--100\,keV} spectral shape softened during the off states compared to the average spectral shape{\oldbf , which is likely driven by a reduction in the mass accretion rate.  While} the {\oldbf duration} of the first four off states in Vela X-1 is similar to our first \xmm\ observation of 4U 1909$+$07{\oldbf , it is important to note that the \integral\ observations do not cover energies below 10\,keV, where effects due to absorption may be present and the observed phenomenology could not be directly compared with our \xmm\ observations.}

To place constraints on the possible accretion geometry of 4U 1909$+$07 during the off state, we compare our \xmm\ campaign with \nustar\ and \xmm\ observations of Vela X-1.  In their \nustar\ campaign of Vela X-1, which consisted of two observations that were performed at similar orbital phases, \citet{2014ApJ...780..133F} found a pulse dropout that persisted for $\sim$450\,s, or about $\sim$1.5 pulse cycles.  The {\oldbf 3--5\,keV/20--30\,keV hardness ratio was observed to soften during the off state, which suggests that it is not linked to obscuration effects from the stellar wind of the B0.5 Ib star.  The authors additionally} observed {\oldbf that} the power law photon index {\oldbf steepens during the offstate} and {\oldbf the} high-energy cutoff {\oldbf present in the average spectrum} was no longer measurable{\oldbf , supporting the premise that the offstate is not tied to absorption}.  Similar off states were also reported in \citet{2022A&A...660A..19D} {\oldbf also} using \nustar.  While we cannot observe changes in the high-energy cutoff due to the limited bandpass of \xmm, we found that the spectral shape in 4U 1909$+$07 during the off state is softer compared to both our time-averaged and time-resolved results (see Figure~\ref{4U 1909 Figure 8}(d)).  This spectral softening, which is coupled with no significant change in the intrinsic neutral hydrogen column density, may be evidence of a reduced mass accretion rate \citep[e.g. Vela X-1; GX 301-2,][]{2008A&A...492..511K,2011A&A...525L...6G}.

\subsection{Pulse Dropout Mechanisms}

{\oldbf \subsubsection{Reprocessing in the Stellar Wind}}
We first consider the possibility that the observed pulse dropout in 4U 1909$+$07 may be solely driven by {\oldbf scattering} effects due to the dense wind of the B0-3 I star.  {\oldbf For an X-ray pulsar embedded in optically thick absorbing material, the strength of its pulsations may be significantly reduced due to different light travel delays in the scattering medium.  In the case of significant reprocessing, the pulsar beam may be completely scattered out of the line of sight, resulting in a turn off of observable pulsations.}  If the pulse dropout is entirely dependent on scattering effects due to the stellar wind, the dropout and subsequent turn on of pulsations may be linked with no significant change in the intrinsic X-ray flux {\oldbf \citep[see OAO 1657-415;][]{2014MNRAS.442.2691P}}.  Our \xmm\ observations{\oldbf ,} however, show that the pulse dropout in 4U 1909$+$07 is coincident with a large and rapid reduction in the 1--10\,keV flux on a timescale of a single pulse period (see Figure~\ref{4U 1909 Figure 5}), providing evidence against this model.  Since low-energy photons are more easily scattered out of the line of sight compared to hard X-ray photons, we additionally would expect the spectral shape to be harder during the pulse dropout.  Marginal spectral softening was observed with \xmm\, which further suggests that the weakening of pulsations is not driven by absorption.

{\oldbf It is also worth noting, that for slow pulsars rotating on the order of a few hundred seconds such as that observed in 4U 1909$+$07, pulsations may still be visible in the case they are not completely scattered by the optically thick material in the wind \citep{2015MNRAS.454.4467P}.  To determine if pulsations may still be detected, we estimate the decoherence timescale using Equation A.9 in \citet{1997ApJ...474..414C}.  This depends on the orbital period, neutron star rotation period and the projected semi-major axis of the source.  This is found to be on the order of $\sim$60\,hr, which is significantly longer than the $\sim$602.62\,s duration of the observed dropout.}

{\oldbf \subsubsection{Reduced Mass Accretion Rate}}
\label{Propeller Effect Section}
An alternative possibility is that the observed pulse dropout is linked to a large reduction in the mass accretion rate, resulting in the ``propeller effect'' where the ram pressure due to accretion no longer overcomes the pressure due to the magnetic field of the neutron star \citep{1975A&A....39..185I}.  This has been observed in the High-mass X-ray binary pulsars V 0332$+$53 and 4U 0115$+$63 \citep{2016A&A...593A..16T}, SMC X-2 \citep{2017ApJ...834..209L} and EXO 2030$+$375 \citep{2017A&A...606A..89F,2021JApA...42...33J}, where at very low accretion rates the centrifugal barrier driven by the rotating magnetosphere of the neutron star is sufficient enough to stop the flow of accreting material.  If the ``propeller effect'' is the cause of the observed pulse dropout, we would expect to simultaneously observe a large decrease in the unabsorbed flux along with a spectral softening that may be tied to a reduced efficiency of bulk Comptonization{\oldbf ,} as well as residual emission from the polar region of the neutron star.  In our \xmm\ observations, we measure the {\oldbf 1--10\,keV} unabsorbed flux to be (2.1$\pm$0.4)$\times$10$^{-11}$\,erg cm$^{-2}$ s$^{-1}$, {\oldbf which corresponds to an unabsorbed 1--10\,keV luminosity of (8$^{+2}_{-1}$)$\times$10$^{34}$\,erg s$^{-1}$ assuming a distance of $\sim$4.85\,kpc {\oldbf \citep{2015A&A...578A.107M}}.  This is only $\sim$20$\%$ of} the average flux (see Section~\ref{sec:ks-integrated}).  We {\oldbf also} find that the spectral shape is softer in the interval from immediately preceding the dip to just after the dip, which agree with predictions supported by the ``propeller effect''.

While the observed drop in the X-ray flux and spectral softening are consistent with predictions linked to the ``propeller effect'', it is important to note that the duration of the observed pulse dropout was a single $\sim$602\,s pulse period and that the unabsorbed X-ray flux increased by a factor of $\sim$5 on a timescale of a few kiloseconds.  The expected duration of magnetic gating mechanisms linked to the propeller effect is expected to persist on much longer timescales \citep[see e.g.][and references therein]{2011A&A...525L...6G}, which is evidence that even if the pulse dropout is driven by the propeller effect{\oldbf , then} that additional mechanisms must be at work to account for its short timescale.

{\oldbf One way to account for the observed dropout persisting one pulse period is if the propeller effect is accompanied} by low-density regions in the wind of the B0-3 I supergiant star, which also results in a reduced accretion rate.  In their hydrodynamic simulations of the inhomogeneous winds of OB supergiant stars in SGXBs, \citet{2015A&A...575A..58M} found that the expansion of bow shocks due to the interaction between the stellar wind and the neutron star results in a large decrease in accretion.  This drives off states where pulsations are no longer detected.  Additionally, regions of increased density are found in the wind resulting in rapid flaring activity and an increase in the pulsed fraction of the observed emission.  Both flaring behavior and rapid off states with durations on the order of a few hundred seconds were observed in the SGXBs Vela X-1 and GX 301-2 \citep{2008A&A...492..511K,2011A&A...535A...9F}, which is consistent with variations in the wind density.
  
{\oldbf If a low-density cavity is present in the stellar wind, a reduction in the neutral hydrogen column density may be observed.  In our pulse-to-pulse spectral analysis (see Section~\ref{Pulse-to-Pulse Spectral Analysis}), we found that $N_{\rm H}$ decreases by a factor of five, which is consistent with a significant decrease in the mass accretion rate due to low density regions such as cavities in the wind.  Additionally, flares that repeat on a timescale of about $\sim$2--2.5\,ks were observed in the first observation (see Figure~\ref{4U 1909 Figure 5}(b) and (c)), which can naturally be explained by alternating high and low density regions in the stellar wind of the B0-3 I star.}

{\oldbf An alternative mechanism that may result in the observed duration of the pulse dropout is if the reduced mass accretion rate is driven by the quasi-spherical settling accretion regime \citep{2012MNRAS.420..216S,2013MNRAS.428..670S}.  This has been applied to sources like the bright symbiotic X-ray binary GX 1$+$4, which shows a steady spin-down torque with respect to luminosity on long timescales \citep{2012A&A...537A..66G} as well as Classical SGXBs.  In X-ray binaries hosting slowly rotating pulsars that accrete at low to moderate X-ray luminosities, the accreted material may not efficiently cool down due to Compton processes. In this case, the gas cannot penetrate the magnetosphere of the neutron star via Rayleigh-Taylor instabilities and it instead accumulates in a hot quasi-spherical shell above the neutron star magnetosphere \citep{1981MNRAS.196..209D}.  The collapse of the quasi-spherical shell due to inefficient radiative cooling processes may drive off states as well as flaring activity that have been observed in Classical SGXBs as well as in Supergiant Fast X-ray Transients \citep[e.g. IGR J11215-5952,][]{2017ApJ...838..133S}.

In our \xmm\ campaign of 4U 1909$+$07, we observe that the off state, where pulsations of the $\sim$602\,s neutron star were no longer detected, preceded a large-scale X-ray flare where the unabsorbed 1--10\,keV dramatically increased on a timescale of minutes (see Figure~\ref{4U 1909 Figure 6}).  This flux dropout and subsequent flaring behavior may be similar to what is observed in the classical SGXBs Vela X-1 and 4U 1907$+$09 \citep{2008A&A...492..511K,1997ApJ...479L..47I}, where a drop in the accretion may possibly be explained by the collapse of a hot quasi-spherical shell of matter onto the neutron star.}

{\oldbf \subsubsection{Constraints on the Settling Accretion Regime}
\label{LowRegime}

To investigate whether the quasi-spherical settling accretion regime can be applied to our \xmm\ observations of 4U 1909$+$07, we compared the characteristic timescales for plasma cooling as outlined in \citet{2012MNRAS.420..216S}. At high luminosities, where material is sufficiently cooled by Compton processes, the timescale where material is accreted onto the neutron star equals the free-fall timescale

\begin{equation}
t_{\mathrm{ff}}(R) =\left(\frac{R^3}{2 G M_{\mathrm{NS}}}\right)^{\frac{1}{2}},
\end{equation}

\noindent
where the height of the accreted plasma, $R_{\rm A}$, equals the Alfv\'en radius where a possible accretion disk is expected to be truncated.  The Alfv\'en radius is proportional to the magnetic dipole moment of the magnetic field of the neutron star, the mass of the neutron star{\oldbf ,} and mass accretion rate.  The magnetic dipole moment can be expressed in terms of the magnetic field of the surface of the neutron star, the radius of the neutron star and the distance from the surface of the neutron star.  Since the X-ray luminosity is directly proportional to the mass accretion rate, we can express the Alfv\'en radius in terms of the magnetic dipole moment of the magnetic field of the neutron star, the mass of the neutron star, X-ray luminosity and the efficiency in converting the accreted material into X-ray radiation, $\Lambda$, which depends on the accretion geometry.  We note that while $\Lambda$ is difficult to estimate, it can be approximated in the disk application as
\begin{equation}
  \Lambda \sim 0.22 \alpha^{18/69},
\end{equation}

\noindent
where $\alpha$ denotes the Shakura-Sunyaev parameter.  A reasonable approximation is to set $\Lambda$=0.5, which is used in \citet{2015MNRAS.454.2714M}{\oldbf , introducing} a systematic uncertainty due to the unknown accretion geometry.

It is important to note that the flux used to calculate the Alfv\'en radius is the bolometric X-ray flux, and, therefore, a bolometric correction from the 1--10\,keV unabsorbed flux observed using \xmm\ must be applied.  From archival contemporaneous observations of 4U 1909$+$07 with \nustar\ and \swift\ \citep{2020MNRAS.498.4830J,2023ApJ...948...45I}, we calculated the ratio of the unabsorbed 0.5--100\,keV and 0.5--10\,keV fluxes using the \texttt{highecut} model is in the range 2.12--2.15.  We, therefore, used the averaged value of 2.13 and estimated the bolometric X-ray flux {\oldbf and luminosity} during the off state to be (4.5$^{+1.0}_{-0.8}$)$\times$10$^{-11}$\,erg cm$^{-2}$ s$^{-1}$ {\oldbf and (1.8$^{+0.4}_{-0.3}$)$\times$10$^{35}$\,erg s$^{-1}$ assuming a distance of $\sim$4.85\,kpc \citep{2015A&A...578A.107M}.  The calculated bolometric luminosity for 4U 1990$+$07 is {\oldbf comparable within the uncertainties} compared to other X-ray pulsars where pulsations were continued to be observed, such as the Be X-ray binary EXO 2030$+$375 where the pulsations were detected at 0.5--30\,keV luminosities as low as $\sim$2.5$\times$10$^{35}$\,erg s$^{-1}$ \citep{2021JApA...42...33J}.}

In Figure~\ref{4U 1909 Figure 9}(a), we show the free-fall timescales of the accreted plasma for magnetic fields between 10$^{11}$\,G and 10$^{13}$\,G, which are reasonable field strengths for an accretion-powered pulsar.  We find the free-fall timescales at a bolometric X-ray luminosity 1.8$^{+0.4}_{-0.3}\times$10$^{35}$\,erg s$^{-1}$ to be between 0.4--20\,seconds for magnetic fields between 10$^{11}$\,G and 10$^{13}$\,G, which are indicated by the red and blue dashed lines in Figure~\ref{4U 1909 Figure 9}(a), respectively.

\begin{figure*}[ht]
    \centering
    \begin{tabular}{c}
    \subfigure
    {
        \includegraphics[width=0.6\textwidth]{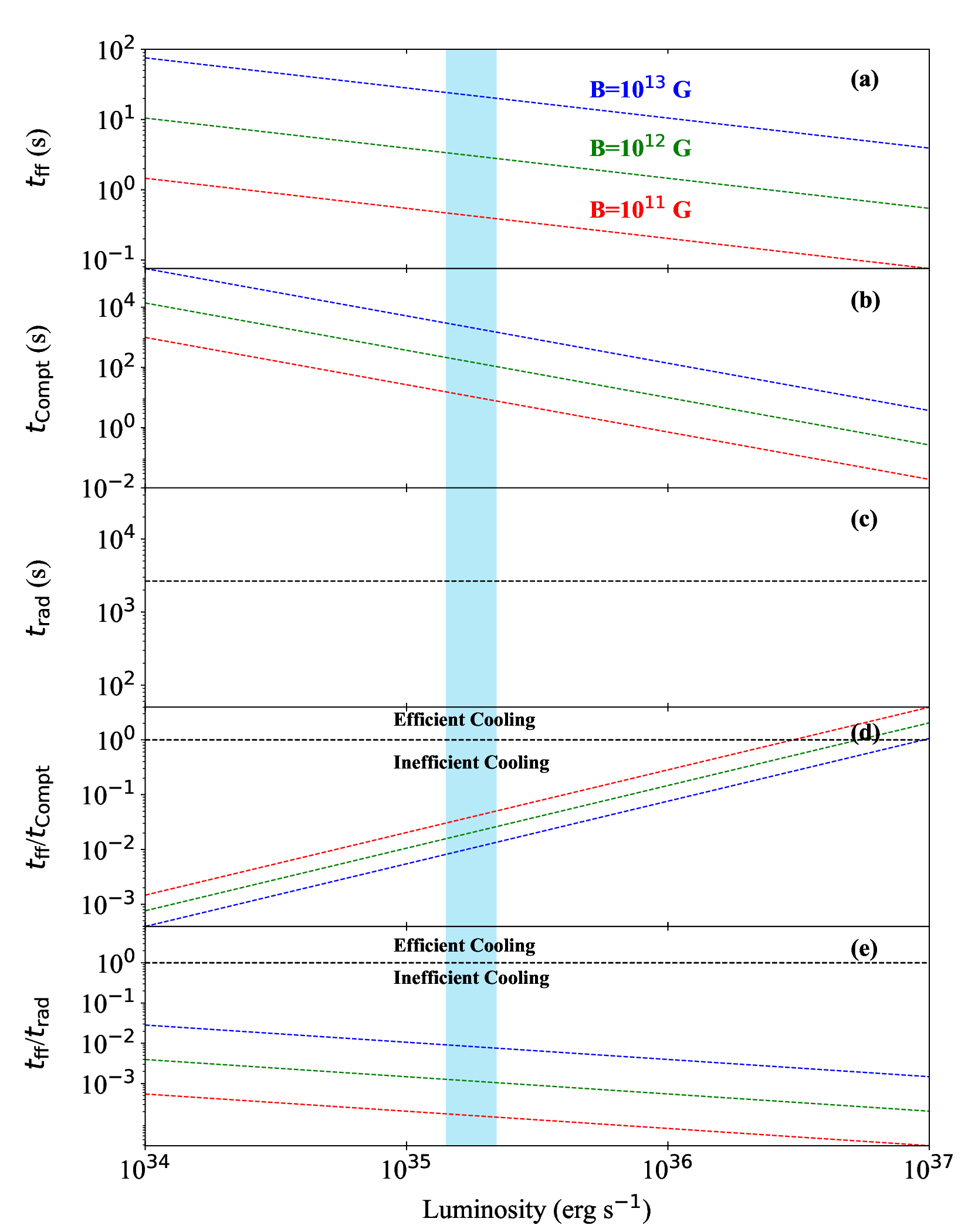}
        \label{Cehck Strength}
    }
    \end{tabular}
    \caption{The (a) free-fall, (b) Compton and (c) radiative cooling timescales of the accretion flow as a function of X-ray luminosity.  The ratios of the free-fall to Compton and free-fall to radiative cooling timescales are shown in panels (d) and (e).  The luminosity of the off-state is indicated by the cyan shaded region.  The red, green and blue dashed lines indicate magnetic field strengths of 10$^{11}$\,G, 10$^{12}$\,G and 10$^{13}$\,G, respectively (see text for details).}
    \label{4U 1909 Figure 9}
\end{figure*}

At low accretion rates, the plasma cooling efficiency is significantly reduced, which is governed by the Compton cooling and radiative cooling timescales, respectively (see Figure~\ref{4U 1909 Figure 9}(b)--(c)).  The Compton cooling time,

\begin{equation}
  \label{Compton cooling}
t_{\mathrm{Comp}} = \frac{3 k T}{4 \sigma_{\mathrm{T}} c U_{\mathrm{ph}}},
\end{equation}

\noindent
is driven by the timescale in which electrons lose energy due to the inverse Compton process in the radiation field near the neutron star.  This is proportional to the Boltzmann constant, the temperature of the plasma, the Thomson cross section, speed of light{\oldbf ,} and the energy density of the radiation field

\begin{equation}
U_{\mathrm{ph}} = \frac{L_{\mathrm{X}}}{4 \pi R^2 c},
\end{equation}

\noindent
which scales directly with X-ray luminosity and inversely with the height of the accreted plasma, which we set to the Alfv\'en radius.  Substituting the energy density of the photon field into Equation~\ref{Compton cooling}, the Compton cooling timescale can now be expressed in terms of the Alfv\'en radius and bolometric X-ray luminosity.  During the observed off-state, the Compton cooling timescale is found to be on the order of 10$^{2}$--10$^{3}$\,seconds, which is significantly longer than the free-fall timescale (see Figure~\ref{4U 1909 Figure 9}(d)). This indicates that accretion proceeds in a radiatively inefficient regime.

At this low X-ray luminosity, radiative cooling where the energy loss proceeds via thermal bremsstrahlung emission

\begin{equation}
t_{\rm Brems} = \frac{3 k_{\rm B} T}{2 \mu_m n \Lambda(T)},
\end{equation}

\noindent
which depends on is the Boltzmann constant, plasma temperature, the particle number density, and the cooling function $\Lambda\left(T\right)$.  At a plasma temperature of 10\,keV, the cooling function can be expressed using Equation 62 in \citet{2012MNRAS.420..216S}, which depends directly on the square root of temperature

\begin{equation}
  \Lambda\left(T\right)=2.5\times10^{-27} T^{1/2}.
\end{equation}

We find this timescale to be $\sim$2.6$\times$10$^{3}$\,seconds, further suggesting that accretion proceeds in the radiatively inefficient regime (see Figure~\ref{4U 1909 Figure 9}(e)) and that the quasi-spherical settling accretion regime may explain the observed pulse dropout and subsequent recovery on the free-fall timescale once efficient cooling resumes.}

\subsection{Constraints on the Magnetic Field}

In X-ray binaries that host accretion powered pulsars, the magnetic field of the neutron star can be directly determined by the energy of cyclotron resonant scattering features (CRSFs), which form when X-ray photons scatter off electrons that are quantized onto Landau-levels in the strong magnetic field of the neutron star \citep{2007A&A...472..353S}.  We note while \citet{2013ApJ...779...54J} observed a possible CRSF at 44\,keV using \suzaku\ HXD-PIN data, suggesting a magnetic field on the order of $\sim$3.8$\times$10$^{12}$\,G, no CRSFs were found in the broadband X-ray spectra of 4U 1909$+$07 using \nustar\ and \swift\ data \citep{2020MNRAS.498.4830J,2023ApJ...948...45I}.  This may suggest the magnetic field is seen under small viewing angles \citep{2007A&A...472..353S} or that the magnetic field is either too small or too large to produce an observable CRSF in the 3--79\,keV band.

If the observed dropout of the $\sim$602\,s pulsation cycle (see Figure~\ref{4U 1909 Figure 5}) was driven by the propeller effect (see Section~\ref{Propeller Effect Section}), we can instead indirectly place a constraint on the magnetic field of the neutron star assuming that the Alfv\'en radius (see Section~\ref{LowRegime}) is equal to the corotation radius where the angular velocity of the accreted material is equal to the corotating Keplerian velocity.   Since the propeller regime is defined where the corotation radius and the Alfv\'en radius are equal, the magnetic field of the neutron star can be expressed as

\begin{equation}
\label{Propeller Magnetic Field}
\begin{split}
  & B = \left(4.8 \times 10^{10}\right) P^{7/6} \left(\frac{\Lambda}{0.1}\right)^{-7/4} \left(\frac{F_{\rm X}}{1.0\times10^{-9}\,{\rm erg}\,{\rm cm}^{-2}\,{\rm s}^{-1}}\right)^{1/2} \\
  & \times \left(d \over 1\,{\rm kpc}\right) \left(\frac{M}{1.4\,M_\odot}\right)^{1/3} \left(R \over 10^{6}\,{\rm cm}\right)^{-5/2} {\rm G},
\end{split}
\end{equation}

\noindent which is directly proportional to the neutron star rotation period, X-ray flux, distance and mass of the neutron star{\oldbf ,} and inversely proportional to neutron star radius and efficiency in converting the accreted material into X-ray radiation, $\Lambda${\oldbf , which we previously set to 0.5 (see Section~\ref{Propeller Effect Section})}.  We note that accretion in 4U 1909$+$07 is likely to be mediated via the wind of the B0-3 I star; however, a transient prograde accretion disk is possible since the neutron star has been observed to spin up over the past 20\,years (see Figure~\ref{4U 1909 Figure 2}).

Assuming a pulse period of $\sim$602.62\,s, a bolometric X-ray luminosity of {\oldbf 1.8$^{+0.4}_{-0.3}\times$10$^{35}$\,erg s$^{-1}$} (see Section~\ref{LowRegime}), neutron star mass of $\sim$1.4\,$M_\odot$, and a neutron star radius of 12\,km \citep[see e.g.][]{2021PhRvD.104f3003L}, we calculate the magnetic field to be {\oldbf $\sim$7.4$\times$10$^{12}$\,G}.  Similar measurements have been used to calculate the magnetic fields of neutron stars in BeXBs \citep[see e.g. 4U 0115+63, SMC X-2;][]{2016A&A...593A..16T,2017ApJ...834..209L} and ultraluminous X-ray pulsars \citep{2016MNRAS.457.1101T} and have been compared with those derived by known CRSFs.

{\oldbf We can alternatively place constraints on the magnetic field of the neutron star by assuming accretion proceeds in the quasi-spherical settling accretion regime.  In this case, the neutron star spin period is regulated by an equilibrium between the accretion torque supplied by the captured stellar wind{\oldbf ,} and the angular momentum transport through a hot, quasi-static shell above the magnetosphere. The corresponding equilibrium spin period can be expressed as

\begin{equation}
  \label{Quasi-Spherical Magnetic Field}
  \begin{split}
    & P_{\rm eq}^{*}=1300 \left(\frac{\mu}{10^{30}\,G cm^{3}}\right)^{12/11}\left(\frac{P_{\rm orb}}{10\,{\rm d}}\right)\left(\frac{\dot{M}}{10^{16}\,{\rm g}\,{\rm s}^{-1}}\right)^{-4/11} \\
    & \times\left(\frac{v}{1000\,{\rm km}\,{\rm s}^{-1}}\right)^{4},
  \end{split}
\end{equation}
\noindent
which depends on the magnetic dipole moment $\mu$, the orbital period, mass accretion rate and wind velocity.  For our calculation, we assume the equilibrium period is the $\sim$602.62\,s neutron star rotation period and the orbital period is $\sim$4.4\,days.  We calculate the mass accretion rate using the bolometric luminosity, where we similarly apply a bolometric correction of $\sim$2.13 to the average X-ray luminosity reported in Table~\ref{Time-Averaged Spectral Parameters}.  Additionally, we similarly set $\Lambda=$0.5, which reasonably approximates the efficiency in converting the accreted material into X-ray radiation.

It is important to note that the calculated magnetic field strongly depends on the velocity of the wind of the donor star, which is typically between 500--2000\,km s$^{-1}$.  In the case that the stellar wind velocity is 1000\,km s$^{-1}$, we find the magnetic field of the neutron star to be $\sim$1.2$\times$10$^{12}$\,G, which is in the typical range for accretion-powered pulsars.  Due to the large dependence on the velocity of the wind of the B0--3 I star, this can be up to an order of magnitude larger for slower wind speeds of 500\,km s$^{-1}$.}

\section{Summary and Conclusions}
\label{Conclusion}

In this paper, we have presented two \xmm\ observations of 4U 1909$+$07, which were performed at similar orbital phases and spanning one cycle of its $\sim$15.2\,day superorbital modulation.  Comparing our timing results with the long-term spin-frequency evolution of 4U 1909$+$07, we found that the neutron star continues to show a global spin-up trend.  We measured the spin period derivative to be (-3.09\,$\pm$\,0.07)$\times$10$^{-9}$\,s\,s$^{-1}$, which is considerably smaller than that reported in \citet{2023ApJ...948...45I}.  The shape of the pulse profiles show no significant energy dependence in the 1--10\,keV band, which is comparable to earlier results using \rxte\, \suzaku\ XIS, \astrosat\ and \nustar.  This is not surprising since broadband timing shows the shape of the pulse profiles changes considerably above 10\,keV, which is not covered by \xmm.  The similar pulse profile shape compared to archival measurements is suggestive of a persistent emission geometry of the pulsar as a function of time.

From our timing results, we detected a possible dropout of the $\sim$602\,s pulsation cycle for the first time.  The possible pulse dropout is coincident with a significant decrease in the 1--10\,keV {\oldbf luminosity}, which we found to be {\oldbf (8$^{+2}_{-1}$)$\times$10$^{34}$\,erg s$^{-1}$} with EPIC-pn  {\oldbf assuming a distance of 4.85\,kpc} {\oldbf \citep{2015A&A...578A.107M}}.  We measured the duration of the pulse dropout to be on the order of $\sim$602\,s, which is the duration of a single pulse.  We additionally found marginal evidence of spectral softening during the off state, which is similar to GX 301-2 and Vela X-1, albeit a weaker trend.  The coincidence between the pulse and flux dropout and spectral softening may be evidence that 4U 1909$+$07 briefly went into the propeller regime, {\oldbf possibly} due to a low-density cavity in the wind of the B0--B3 I supergiant star.

We note that while no CRSF was found in earlier broadband X-ray analyses, if the pulse dropout is indeed tied to 4U 1909$+$07 briefly entering the propeller regime, we can begin to place constraints on the magnetic field of the $\sim$602\,s neutron star.  Assuming that the {\oldbf Alfv\'en} radius is equal to the corotation radius, we measured the magnetic field of the X-ray pulsar to be {\oldbf $\sim$7.4$\times$10$^{12}$\,G}.  {\oldbf Alternative constraints on the magnetic field of the X-ray pulsar can be placed if the reduction in the mass accretion rate is instead tied to the quasi-spherical settling accretion regime.  If the neutron star is rotating near equilibrium, we measured its magnetic field to be $\sim$1.2$\times$10$^{12}$\,G.}  We note that {\oldbf these measurements are} hampered due to uncertainties from the still unknown accretion geometry.  Alternatively, the non-detection of a CRSF may instead suggest the strong magnetic field of the neutron star is seen under small viewing angles.

While we detected strong Fe K$\alpha$ and Fe K$\beta$ emission features in the \xmm\ spectra, we found no evidence of a Compton shoulder on the red wing of the 6.4\,keV emission feature.  The Compton shoulder was previously observed in a \chandra\ observation of 4U 1909$+$07, albeit at a much lower neutral hydrogen column density compared to our \xmm\ spectra.  While this is consistent with smearing effects {\oldbf that are likely} due to second-order Compton scattering, we cannot confirm or refute the possibility of a Compton shoulder due to the lower spectral resolution of EPIC-pn compared to \chandra\ HETG.

To further probe the environment around the SGXB 4U 1909$+$07, much higher resolution soft X-ray spectroscopy is required.  The study would benefit from \xrism\ observations of 4U 1909$+$07, which may be able to detect a weak Compton shoulder if it indeed is present.  Additionally, the study of 4U 1909$+$07 would benefit from hydrodynamical modeling similar to that done for Vela X-1 \citep{2015A&A...575A..58M}{\oldbf ,} as well as additional timing measurements to confirm or refute the possibility that the observed pulse dropout is due to low-density cavities in the wind.

\acknowledgements

{\oldbf We thank the anonymous referee for useful comments that helped improve the quality of the manuscript.}  We {\oldbf also} thank Drs. Koji Mukai, {\oldbf Patrizia Romano and Enrico Bozzo} for useful discussions{\oldbf , and the \newtonshort\ Science Operations Centre for their support in scheduling and the execution of these observations.}  The material is based upon work supported by NASA under award number 80GSFC21M0006.  This research has made use of the \xmm\ Science Analysis System (XMMSAS) developed by ESA.  We thank the \xmm\ Guest Observer Grant 80NSSC22K0801 for support.


\begin{thebibliography}{}

\bibitem[Bailer-Jones et al.(2021)]{2021AJ....161..147B} Bailer-Jones, C.~A.~L., Rybizki, J., Fouesneau, M., et al.\ 2021, \aj, 161, 147. doi:10.3847/1538-3881/abd806

\bibitem[Ballhausen et al.(2017)]{2017A&A...608A.105B} Ballhausen, R., Pottschmidt, K., F{\"u}rst, F., et al.\ 2017, \aap, 608, A105. doi:10.1051/0004-6361/201730845

\bibitem[Ballhausen et al.(2024)]{2024A&A...688A.214B} Ballhausen, R., Thalhammer, P., Pradhan, P., et al.\ 2024, \aap, 688, A214. doi:10.1051/0004-6361/202348595

\bibitem[Bozzo et al.(2017)]{2017A&A...608A.128B} Bozzo, E., Bernardini, F., Ferrigno, C., et al.\ 2017, \aap, 608, A128. doi:10.1051/0004-6361/201730398

\bibitem[Briel et al.(2005)]{2005SPIE.5898..194B} Briel, U.~G., Burwitz, V., Dennerl, K., et al.\ 2005, \procspie, 5898, 194. doi:10.1117/12.616824

\bibitem[Canizares et al.(2005)]{2005PASP..117.1144C} Canizares, C.~R., Davis, J.~E., Dewey, D., et al.\ 2005, \pasp, 117, 1144. doi:10.1086/432898

\bibitem[Chakrabarty et al.(1997)]{1997ApJ...474..414C} Chakrabarty, D., Bildsten, L., Grunsfeld, J.~M., et al.\ 1997, \apj, Torque Reversal and Spin-down of the Accretion-powered Pulsar 4U 1626-67, 474, 1, 414. doi:10.1086/303445

\bibitem[Corbet \& Krimm(2013)]{2013ATel.5119....1C} Corbet, R.~H.~D. \& Krimm, H.~A.\ 2013, The Astronomer's Telegram, 5119

\bibitem[Corbet 
\& Krimm(2013)]{2013ApJ...778...45C} Corbet, R.~H.~D., \& Krimm, H.~A.\ 2013, \apj, 778, 45

\bibitem[Davies \& Pringle(1981)]{1981MNRAS.196..209D} Davies, R.~E. \& Pringle, J.~E.\ 1981, \mnras, 196, 209. doi:10.1093/mnras/196.2.209

\bibitem[de Kool \& Anzer(1993)]{1993MNRAS.262..726D} de Kool, M. \& Anzer, U.\ 1993, \mnras, 262, 726. doi:10.1093/mnras/262.3.726

\bibitem[den Herder et al.(2001)]{2001A&A...365L...7D} den Herder, J.~W., Brinkman, A.~C., Kahn, S.~M., et al.\ 2001, \aap, 365, L7. doi:10.1051/0004-6361:20000058

\bibitem[Diez et al.(2022)]{2022A&A...660A..19D} Diez, C.~M., Grinberg, V., F{\"u}rst, F., et al.\ 2022, \aap, 660, A19. doi:10.1051/0004-6361/202141751

\bibitem[Diez et al.(2023)]{2023A&A...674A.147D} Diez, C.~M., Grinberg, V., F{\"u}rst, F., et al.\ 2023, \aap, 674, A147. doi:10.1051/0004-6361/202245708

\bibitem[El Mellah et al.(2019)]{2019A&A...622A.189E} El Mellah, I., Sander, A.~A.~C., Sundqvist, J.~O., et al.\ 2019, \aap, 622, A189. doi:10.1051/0004-6361/201834498

\bibitem[Falanga et al.(2015)]{2015A&A...577A.130F} Falanga, M., Bozzo, E., Lutovinov, A., et al.\ 2015, \aap, Ephemeris, orbital decay, and masses of ten eclipsing high-mass X-ray binaries, 577, A130. doi:10.1051/0004-6361/201425191

\bibitem[Fortin et al.(2023)]{2023A&A...671A.149F} Fortin, F., Garc{\'\i}a, F., Simaz Bunzel, A., et al.\ 2023, \aap, 671, A149. doi:10.1051/0004-6361/202245236

\bibitem[F{\"u}rst et al.(2010)]{2010A&A...519A..37F} F{\"u}rst, F., Kreykenbohm, I., Pottschmidt, K., et al.\ 2010, \aap, 519, A37. doi:10.1051/0004-6361/200913981

\bibitem[F{\"u}rst et al.(2011a)]{2011A&A...535A...9F} F{\"u}rst, F., Suchy, S., Kreykenbohm, I., et al.\ 2011a, \aap, 535, A9. doi:10.1051/0004-6361/201117665

\bibitem[F{\"u}rst et al.(2011b)]{2011A&A...525A..73F} F{\"u}rst, F., Kreykenbohm, I., Suchy, S., et al.\ 2011b, \aap, 525, A73. doi:10.1051/0004-6361/201015636

\bibitem[F{\"u}rst et al.(2012)]{2012A&A...547A...2F} F{\"u}rst, F., Pottschmidt, K., Kreykenbohm, I., et al.\ 2012, \aap, 547, A2. doi:10.1051/0004-6361/201219845

\bibitem[F{\"u}rst et al.(2014)]{2014ApJ...780..133F} F{\"u}rst, F., Pottschmidt, K., Wilms, J., et al.\ 2014, \apj, 780, 133. doi:10.1088/0004-637X/780/2/133

\bibitem[F{\"u}rst et al.(2017)]{2017A&A...606A..89F} F{\"u}rst, F., Kretschmar, P., Kajava, J.~J.~E., et al.\ 2017, \aap, 606, A89. doi:10.1051/0004-6361/201730941

\bibitem[Gaia Collaboration et al.(2023)]{2023A&A...674A...1G} Gaia Collaboration, Vallenari, A., Brown, A.~G.~A., et al.\ 2023, \aap, 674, A1. doi:10.1051/0004-6361/202243940

\bibitem[Giacconi et al.(1974)]{1974ApJS...27...37G} Giacconi, R., Murray, S., Gursky, H., et al.\ 1974, \apjs, 27, 37. doi:10.1086/190288

\bibitem[G{\"o}{\u{g}}{\"u}{\c{s}} et al.(2011)]{2011A&A...525L...6G} G{\"o}{\u{g}}{\"u}{\c{s}}, E., Kreykenbohm, I., \& Belloni, T.~M.\ 2011, \aap, 525, L6. doi:10.1051/0004-6361/201015905

\bibitem[Gonz{\'a}lez-Gal{\'a}n et al.(2012)]{2012A&A...537A..66G} Gonz{\'a}lez-Gal{\'a}n, A., Kuulkers, E., Kretschmar, P., et al.\ 2012, \aap, Spin period evolution of GX 1+4, 537, A66. doi:10.1051/0004-6361/201117893

\bibitem[Grebenev \& Sunyaev(2007)]{2007AstL...33..149G} Grebenev, S.~A. \& Sunyaev, R.~A.\ 2007, Astronomy Letters, 33, 149. doi:10.1134/S1063773707030024

\bibitem[Hemphill et al.(2014)]{2014ApJ...792...14H} Hemphill, P.~B., Rothschild, R.~E., Markowitz, A., et al.\ 2014, \apj, 792, 14. doi:10.1088/0004-637X/792/1/14

\bibitem[HI4PI Collaboration et al.(2016)]{2016A&A...594A.116H} HI4PI Collaboration, Ben Bekhti, N., Flöer, L., et al.\ 2016, \aap, 594, A116. doi:10.1051/0004-6361/201629178

\bibitem[Hiltner et al.(1972)]{1972ApJ...175L..19H} Hiltner, W.~A., Werner, J., \& Osmer, P.\ 1972, \apjl, 175, L19. doi:10.1086/180976

\bibitem[Illarionov \& Sunyaev(1975)]{1975A&A....39..185I} Illarionov, A.~F. \& Sunyaev, R.~A.\ 1975, \aap, 39, 185


\bibitem[in 't Zand et al.(1997)]{1997ApJ...479L..47I} in 't Zand, J.~J.~M., Strohmayer, T.~E., \& Baykal, A.\ 1997, \apjl, 479, L47. doi:10.1086/310570

\bibitem[Islam et al.(2023)]{2023ApJ...948...45I} Islam, N., Corbet, R.~H.~D., Coley, J.~B., et al.\ 2023, \apj, 948, 45. doi:10.3847/1538-4357/acbc19

\bibitem[Jaisawal et al.(2013)]{2013ApJ...779...54J} Jaisawal, G.~K., Naik, S., \& Paul, B.\ 2013, \apj, 779, 54. doi:10.1088/0004-637X/779/1/54

\bibitem[Jaisawal et al.(2020)]{2020MNRAS.498.4830J} Jaisawal, G.~K., Naik, S., Ho, W.~C.~G., et al.\ 2020, \mnras, 498, 4830. doi:10.1093/mnras/staa2604

\bibitem[Jaisawal et al.(2021)]{2021JApA...42...33J} Jaisawal, G.~K., Naik, S., Gupta, S., et al.\ 2021, Journal of Astrophysics and Astronomy, 42, 33. doi:10.1007/s12036-021-09699-2

\bibitem[Jansen et al.(2001)]{2001A&A...365L...1J} Jansen, F., Lumb, D., Altieri, B., et al.\ 2001, \aap, 365, L1. doi:10.1051/0004-6361:20000036

\bibitem[Kaastra \& Bleeker(2016)]{2016A&A...587A.151K} Kaastra, J.~S. \& Bleeker, J.~A.~M.\ 2016, \aap, 587, A151. doi:10.1051/0004-6361/201527395

\bibitem[Kallman et al.(2004)]{2004ApJS..155..675K} Kallman, T.~R., Palmeri, P., Bautista, M.~A., et al.\ 2004, \apjs, 155, 675. doi:10.1086/424039

\bibitem[Kraus et al.(1989)]{1989ESASP.296..433K} Kraus, U., Rebetzky, A., Herold, H., et al.\ 1989, Two Topics in X-Ray Astronomy, Volume 1: X Ray Binaries.~ Volume 2: AGN and the X Ray Background, 296, 

\bibitem[Kretschmar et al.(2014)]{2014EPJWC..6406012K} Kretschmar, P., Marcu, D., K{\"u}hnel, M., et al.\ 2014, European Physical Journal Web of Conferences, 64, 06012. doi:10.1051/epjconf/20136406012

\bibitem[Kretschmar et al.(2017)]{2017IAUS..329..355K} Kretschmar, P., Mart{\'\i}nez-N{\'u}{\~n}ez, S., Bozzo, E., et al.\ 2017, The Lives and Death-Throes of Massive Stars, 329, 355. doi:10.1017/S1743921317002411

\bibitem[Kreykenbohm et al.(2008)]{2008A&A...492..511K} Kreykenbohm, I., Wilms, J., Kretschmar, P., et al.\ 2008, \aap, 492, 511. doi:10.1051/0004-6361:200809956

\bibitem[Kuster et al.(2002)]{2002astro.ph..3213K} Kuster, M., Wilms, J., Staubert, R., et al.\ 2002, astro-ph/0203213. doi:10.48550/arXiv.astro-ph/0203213

\bibitem[Leahy(1987)]{1987A&A...180..275L} Leahy, D.~A.\ 1987, \aap, 180, 275

\bibitem[Legred et al.(2021)]{2021PhRvD.104f3003L} Legred, I., Chatziioannou, K., Essick, R., et al.\ 2021, \prd, 104, 063003. doi:10.1103/PhysRevD.104.063003

\bibitem[Levine et al.(2004)]{2004ApJ...617.1284L} Levine, A.~M., Rappaport, S., Remillard, R., et al.\ 2004, \apj, 617, 1284. doi:10.1086/425567

\bibitem[Lutovinov et al.(2017)]{2017ApJ...834..209L} Lutovinov, A.~A., Tsygankov, S.~S., Krivonos, R.~A., et al.\ 2017, \apj, 834, 209. doi:10.3847/1538-4357/834/2/209

\bibitem[Makishima et al.(1999)]{1999ApJ...525..978M} Makishima, K., Mihara, T., Nagase, F., \& Tanaka, Y.\ 1999, \apj, 525, 978 

\bibitem[Maitra et al.(2012)]{2012MNRAS.420.2307M} Maitra, C., Paul, B., \& Naik, S.\ 2012, \mnras, 420, 2307. doi:10.1111/j.1365-2966.2011.20196.x

\bibitem[Manousakis \& Walter(2015)]{2015A&A...575A..58M} Manousakis, A. \& Walter, R.\ 2015, \aap, 575, A58. doi:10.1051/0004-6361/201321414

\bibitem[Malacaria et al.(2016)]{2016A&A...588A.100M} Malacaria, C., Mihara, T., Santangelo, A., et al.\ 2016, \aap, Probing the stellar wind environment of Vela X-1 with MAXI, 588, A100. doi:10.1051/0004-6361/201527009

\bibitem[Markert et al.(1979)]{1979ApJS...39..573M} Markert, T.~H., Winkler, P.~F., Laird, F.~N., et al.\ 1979, \apjs, 39, 573. doi:10.1086/190587

\bibitem[Mart{\'\i}nez-N{\'u}{\~n}ez et al.(2015)]{2015A&A...578A.107M} Mart{\'\i}nez-N{\'u}{\~n}ez, S., Sander, A., G{\'\i}menez-Garc{\'\i}a, A., et al.\ 2015, \aap, 578, A107. doi:10.1051/0004-6361/201424823

\bibitem[Mart{\'\i}nez-N{\'u}{\~n}ez et al.(2017)]{2017SSRv..212...59M} Mart{\'\i}nez-N{\'u}{\~n}ez, S., Kretschmar, P., Bozzo, E., et al.\ 2017, \ssr, 212, 59. doi:10.1007/s11214-017-0340-1

\bibitem[Massa et al.(2014)]{2014MNRAS.441.2173M} Massa, D., Oskinova, L., Fullerton, A.~W., et al.\ 2014, \mnras, 441, 2173. doi:10.1093/mnras/stu565

\bibitem[Matt(2002)]{2002MNRAS.337..147M} Matt, G.\ 2002, \mnras, 337, 147. doi:10.1046/j.1365-8711.2002.05890.x

\bibitem[McClintock et al.(1976)]{1976ApJ...206L..99M} McClintock, J.~E., Rappaport, S., Joss, P.~C., et al.\ 1976, \apjl, 206, L99. doi:10.1086/182142

\bibitem[Meszaros \& Nagel(1985)]{1985ApJ...299..138M} Meszaros, P., \& Nagel, W.\ 1985, \apj, 299, 138 

\bibitem[Mushtukov et al.(2015)]{2015MNRAS.454.2714M} Mushtukov, A.~A., Tsygankov, S.~S., Serber, A.~V., et al.\ 2015, \mnras, 454, 2714. doi:10.1093/mnras/stv2182

\bibitem[Ogilvie \& Dubus(2001)]{2001MNRAS.320..485O} Ogilvie, G.~I., \& Dubus, G.\ 2001, \mnras, 320, 485

\bibitem[Palmeri et al.(2003)]{2003A&A...410..359P} Palmeri, P., Mendoza, C., Kallman, T.~R., et al.\ 2003, \aap, 410, 359. doi:10.1051/0004-6361:20031262


\bibitem[Pradhan et al.(2014)]{2014MNRAS.442.2691P} Pradhan, P., Maitra, C., Paul, B., et al.\ 2014, \mnras, 442, 2691. doi:10.1093/mnras/stu1034

\bibitem[Pradhan et al.(2015)]{2015MNRAS.454.4467P} Pradhan, P., Paul, B., Paul, B.~C., et al.\ 2015, \mnras, Is 4U 0114+65 an eclipsing HMXB?, 454, 4, 4467. doi:10.1093/mnras/stv2276

\bibitem[Pradhan et al.(2019)]{2019MNRAS.483.5687P} Pradhan, P., Raman, G., \& Paul, B.\ 2019, \mnras, 483, 5687. doi:10.1093/mnras/sty3441

\bibitem[Pradhan et al.(2023)]{2023ApJ...945...51P} Pradhan, P., Ferrigno, C., Paul, B., et al.\ 2023, \apj, 945, 51. doi:10.3847/1538-4357/acb2cb

\bibitem[Pike et al.(2019)]{2019ApJ...875..144P} Pike, S.~N., Harrison, F.~A., Bachetti, M., et al.\ 2019, \apj, 875, 144
  
\bibitem[Protassov et al.(2002)]{2002ApJ...571..545P} Protassov, R., van Dyk, D.~A., Connors, A., Kashyap, V.~L., \& Siemiginowska, A.\ 2002, \apj, 571, 545
  
\bibitem[Pringle(1996)]{1996MNRAS.281..357P} Pringle, J.~E.\ 1996, MNRAS, 281, 357

\bibitem[Puls et al.(2008)]{2008A&ARv..16..209P} Puls, J., Vink, J.~S., \& Najarro, F.\ 2008, \aapr, 16, 209. doi:10.1007/s00159-008-0015-8

\bibitem[Read et al.(2014)]{2014A&A...564A..75R} Read, A.~M., Guainazzi, M., \& Sembay, S.\ 2014, \aap, 564, A75. doi:10.1051/0004-6361/201423422

\bibitem[Romano et al.(2015)]{2015A&A...576L...4R} Romano, P., Bozzo, E., Mangano, V., et al.\ 2015, \aap, 576, L4. doi:10.1051/0004-6361/201525749

\bibitem[Romano et al.(2025)]{2025arXiv250504452R} Romano, P., Cohen, H.~I., Bozzo, E., et al.\ 2025, , Swift/XRT monitoring of the orbital and superorbital modulations in 4U 1909+07, arXiv:2505.04452. doi:10.48550/arXiv.2505.04452

\bibitem[Sch{\"o}nherr et al.(2007)]{2007A&A...472..353S} Sch{\"o}nherr, G., Wilms, J., Kretschmar, P., et al.\ 2007, \aap, 472, 353. doi:10.1051/0004-6361:20077218

\bibitem[Sguera et al.(2020)]{2020ApJ...900...22S} Sguera, V., Tiengo, A., Sidoli, L., et al.\ 2020, \apj, 900, 22. doi:10.3847/1538-4357/abaa3c

\bibitem[Sidoli et al.(2017)]{2017ApJ...838..133S} Sidoli, L., Tiengo, A., Paizis, A., et al.\ 2017, \apj, 838, 133. doi:10.3847/1538-4357/aa671a

\bibitem[Shakura et al.(2012)]{2012MNRAS.420..216S} Shakura, N., Postnov, K., Kochetkova, A., et al.\ 2012, \mnras, 420, 216. doi:10.1111/j.1365-2966.2011.20026.x

\bibitem[Shakura et al.(2013)]{2013MNRAS.428..670S} Shakura, N., Postnov, K., \& Hjalmarsdotter, L.\ 2013, \mnras, 428, 670. doi:10.1093/mnras/sts062

\bibitem[Str{\"u}der et al.(2001)]{2001A&A...365L..18S} Str{\"u}der, L., Briel, U., Dennerl, K., et al.\ 2001, \aap, 365, L18. doi:10.1051/0004-6361:20000066

\bibitem[Thalhammer et al.(2024)]{2024A&A...688A.213T} Thalhammer, P., Ballhausen, R., Sokolova-Lapa, E., et al.\ 2024, \aap, 688, A213. doi:10.1051/0004-6361/202348594

\bibitem[Torrej{\'o}n et al.(2010)]{2010ApJ...715..947T} Torrej{\'o}n, J.~M., Schulz, N.~S., Nowak, M.~A., et al.\ 2010, \apj, 715, 947. doi:10.1088/0004-637X/715/2/947

\bibitem[Tsygankov et al.(2016)]{2016A&A...593A..16T} Tsygankov, S.~S., Lutovinov, A.~A., Doroshenko, V., et al.\ 2016, \aap, 593, A16. doi:10.1051/0004-6361/201628236

\bibitem[Tsygankov et al.(2016)]{2016MNRAS.457.1101T} Tsygankov, S.~S., Mushtukov, A.~A., Suleimanov, V.~F., et al.\ 2016, \mnras, 457, 1101. doi:10.1093/mnras/stw046

\bibitem[Turner et al.(2001)]{2001A&A...365L..27T} Turner, M.~J.~L., Abbey, A., Arnaud, M., et al.\ 2001, \aap, 365, L27. doi:10.1051/0004-6361:20000087

\bibitem[Vaughan et al.(2003)]{2003MNRAS.345.1271V} Vaughan, S., Edelson, R., Warwick, R.~S., \& Uttley, P.\ 2003, \mnras, 345, 1271

\bibitem[Verner et al.(1996)]{1996ApJ...465..487V} Verner, D.~A., Ferland, G.~J., Korista, K.~T., \& Yakovlev, D.~G.\ 1996, \apj, 465, 487

\bibitem[Watanabe et al.(2003)]{2003ApJ...597L..37W} Watanabe, S., Sako, M., Ishida, M., et al.\ 2003, \apjl, 597, L37. doi:10.1086/379735

\bibitem[Warwick et al.(1988)]{1988MNRAS.232..551W} Warwick, R.~S., Norton, A.~J., Turner, M.~J.~L., et al.\ 1988, \mnras, 232, 551. doi:10.1093/mnras/232.3.551

\bibitem[Wen et al.(2000)]{2000ApJ...532.1119W} Wen, L., Remillard, R.~A., \& Bradt, H.~V.\ 2000, \apj, 532, 1119. doi:10.1086/308604

\bibitem[White et al.(1983)]{1983ApJ...270..711W} White, N.~E., Swank, 
J.~H., \& Holt, S.~S.\ 1983, ApJ, 270, 711

\bibitem[Wilms et al.(2000)]{2000ApJ...542..914W} Wilms, J., Allen, A., 
\& McCray, R.\ 2000, ApJ, 542, 914 

\bibitem[Wood et al.(1984)]{1984ApJS...56..507W} Wood, K.~S., Meekins, J.~F., Yentis, D.~J., et al.\ 1984, \apjs, 56, 507. doi:10.1086/190992

\end{thebibliography}
\end{document}